\documentclass[apj]{emulateapj}
\usepackage{graphicx}
\usepackage{amssymb,amsmath}
\usepackage{natbib}
\usepackage[backref,breaklinks,colorlinks,citecolor=blue]{hyperref}
\usepackage[all]{hypcap}

\makeatletter

\newcommand{\Rmnum}[1]{\expandafter\@slowromancap\romannumeral #1@}
\makeatother

\shorttitle{Updated Emission Spectroscopy of Exoplanet HD~189733\lowercase{b}}
\shortauthors{Todorov et al.}
\begin{document}
\title{Updated {\it Spitzer} Emission Spectroscopy of Bright Transiting Hot Jupiter HD~189733\lowercase{b}}
\author{
Kamen O. Todorov\altaffilmark{1,2,5}, 
Drake Deming\altaffilmark{2},
Adam Burrows\altaffilmark{3},
Carl J. Grillmair\altaffilmark{4}
}
\altaffiltext{1}{Department of Astronomy and Astrophysics, The Pennsylvania State University, University Park, PA~16802, USA}
\altaffiltext{2}{Department of Astronomy, University of Maryland at College park, College Park, MD~20742, USA}
\altaffiltext{3}{Department of Astrophysical Sciences, Princeton University, Princeton, NJ~08544, USA}
\altaffiltext{4}{Spitzer Science Center, California Institute of Technology, Mail Stop 220-6, Pasadena,CA~91125, USA}
\altaffiltext{5}{Current address: Institute for Astronomy, ETH Z\"urich, Wolfgang-Pauli-Strasse 27, 8093 Z\"urich, Switzerland, \href{mailto:todorovk@phys.ethz.ch}{\mbox{todorovk@phys.ethz.ch}}}

\begin{abstract}
We analyze all existing  secondary eclipse time series spectroscopy of hot Jupiter HD~189733b
acquired with the now defunct {\it Spitzer}/IRS instrument. We describe the novel approaches we develop to 
remove the systematic effects and extract accurate secondary eclipse depths as a function of wavelength in order to 
construct the emission spectrum of the exoplanet. We compare our results to a previous study by 
Grillmair et al. that did not examine all data sets available to us. We are able 
to confirm the detection of a water feature near $6\,\mu$m claimed by Grillmair et al. 
We compare the planetary emission spectrum to three model families -- based on isothermal 
atmosphere, gray atmosphere, and two realizations of the complex radiative transfer model by Burrows et al., 
adopted in Grillmair et al.'s study. While we are able to reject the simple 
isothermal and gray models based on the data at the 97\% level just from the IRS data, these rejections hinge on 
eclipses measured within relatively narrow wavelength range, between 5.5 and $7\,\mu$m. 
This underscores the need for observational studies with broad wavelength coverage and 
high spectral resolution, in order to obtain robust information on exoplanet atmospheres.
\end{abstract}



\section{Introduction}
\label{sec:intro}
Over 1820 planets have been discovered so far in orbit around stars 
other than the Sun\footnote{Jean Schneider, \href{http://exoplanet.eu}{exoplanet.eu}, as of 
Sept. 1, 2014.}. A sub-set of these exoplanets often referred to as
``hot Jupiters'', due to their sizes and orbital periods of 
less than 10 days, is particularly amenable to atmospheric
studies via transit observations. 
Secondary eclipse broadband photometry of hot Jupiters has been very successful and planetary 
emission has been measured via this method in numerous investigations 
\citep[e.g.,][]{cha05,dem05,knu08,ste10,dem11,lew13,tod13}.
Secondary eclipse {\it spectroscopy} is much more difficult due to 
photon limits -- the narrow-band secondary eclipse measurements have much lower
signal-to-noise than broadband photometry for a given target. The mid-infrared eclipse depth even for
hot Jupiters is often below $\sim0.5$\%, making this technique 
only possible, with current technology, for transiting planetary systems such as HD~189733 and 
HD~209458, where the host star has high apparent brightness and the
planet-star contrast is also relatively high, leading to measurements with 
relatively high signal-to-noise ratios. Secondary eclipse spectroscopy 
studies have so far been limited to observations in the mid-infrared with 
the {\it Spitzer Space Telescope} \citep{gri07,ric07,gri08,swa08a}, and at shorter
wavelengths with the {\it Hubble Space Telescope} 
\citep[e.g.,][]{swa09a, eva13} and some premier ground-based facilities
\citep[e.g.,][]{cro12}.

The secondary eclipse depth of a planet at a given wavelength is a 
measure of its emission; it is equivalent to the contrast between the planet and 
the star. Therefore, the eclipse depth as a function of wavelength can be used 
to study the planet's emission spectrum, and therefore characterize its atmosphere. 
Investigations utilizing broadband secondary eclipse photometry have 
suggested that hot Jupiters have two classes of atmospheres -- 
with and without a temperature inversion \citep[e.g.,][]{knu08,mac09,tod10}.
A planet with a ``temperature inversion'' is understood to have a layer in its
upper atmosphere that is warmer than the layers below, while in non-inverted
atmospheres the temperature decreases monotonically with altitude.
Some examples of planets with evidence for inverted atmospheres include 
HD~209458b \citep{knu08}, CoRoT-1b \citep{dem11}, XO-4b \citep{tod12}, etc., 
while, e.g., HD~189733b \citep{cha08}, TrES-1 \citep{cha05} and WASP-4b \citep{bee11}, 
have evidence for a lack of temperature inversion in their atmospheres. 
In addition to these two classes, there are several hot Jupiters that have 
ambiguous mid-infrared photometry measurements, compatible with both inverted
and non-inverted models. And in some cases, e.g. CoRoT-2b 
\citep{dem11} and HAT-P-12b \citep{tod13}, the current models both with and 
without a temperature inversion fail to describe the observations completely, possibly 
due to incorrect or incomplete assumptions about the chemical and physical properties of the 
modeled atmospheres. 

The causes for the presence or absence of temperature inversions are currently  
not well-understood. A possible explanation is that planets with inverted 
atmospheres have an additional chemical species in their upper layers that 
causes strong absorption at pressures below $\sim0.01$\,bar, which leads to
extra heating of these atmospheric layers \citep{bur08, for08}. 
The absorber has been proposed to be gas-phase TiO 
\citep{hub03,bur07,bur08,for06,for08}, but this is under debate 
\citep[e.g.,][]{zah09, knu10, par13}. 

Emission spectroscopy of planets can also be used to study the 
climates of hot Jupiters. Since hot Jupiters have very short orbital periods, 
they are expected to synchronize their rotation period with their orbital 
periods within $\sim1$\,Gyr, assuming zero orbital eccentricity 
\citep[e.g.,][]{jac08,cor10}, and hence have permanent day and night sides. 
The transport of heat from the day side 
to the night side of the planet is directly affected by the strength and 
direction of the atmospheric currents. More efficient energy transfer 
leads to a cooler day side. By measuring the dayside thermal emission 
via secondary eclipse observations, we can probe
the heat transport efficiency. This parameter is, however, degenerate 
with the Bond albedo of the planet, which can also cool the day side by reflecting the stellar flux. 
Constraints can be placed on the combination of both parameters by comparison 
to models \citep{cow11}.

The emission spectrum of a planet could also be used to detect and 
measure the abundances of various molecular species in the planetary 
atmosphere. However, most secondary eclipse measurements are based on 
broadband photometry and the ``spectra'' constructed from such observations in  
multiple wavelengths have extremely low resolution. Despite this, there have been 
efforts to evaluate of the abundances of molecules like 
CO$_2$, CO, CH$_4$ and H$_2$O based on broad band photometry
\citep[e.g.,][]{ste10}. However, more recent retrieval efforts
\citep[e.g.,][]{lee12, lin12, lin13, bar13} have suggested that much 
higher spectral resolution and large wavelength coverage are often 
required to reliably measure molecular abundances, especially when considering 
other free parameters like the temperature-pressure structure 
of the atmosphere and the efficiency of heat transport to the night side. 

A previous study with the InfraRed Spectrograph \citep[IRS, ][]{hou04} on 
{\it Spitzer} by \citet{gri08} analyzed 10 time-series emission spectroscopy data sets
on HD~189733b with coverage between $\sim5$ and $\sim14$\,$\mu$m and found a strong 
downturn in planetary emission below $10$\,$\mu$m as well as a spectral feature
associated with water vapor absorption\footnote{Water detections have also been 
reported in the infrared transmission 
spectrum of the planet \citep{tin07, swa08b, bir13}.}.
In addition to these 10 secondary eclipse data sets, 
there are eight in the same wavelength range and four
between 21\,$\mu$m and 40\,$\mu$m that were also observed 
but have never been analyzed. 

While there are dozens of planets observed via {\it Spitzer} 
photometry at the 3.6 and 4.5\,$\mu$m bands during eclipse, only 
two planets have ever been observed via secondary eclipse time-series 
spectroscopy with {\it Spitzer} -- HD~189733b \citep{gri07,gri08} 
and HD~209458b \citep{ric07,swa08a}. This number
will not increase, since after the cryogen on board ran out in 2009 
{\it Spitzer} no longer has spectroscopic capabilities. 
The relatively high temperature of the planet and the relatively low temperature of the star 
result in deep eclipses in an IR-bright target, thus
allowing for relatively high signal-to-noise measurements of the eclipse depths as 
a function of wavelength. (We list the planetary system parameters in \autoref{tab:prop}.)
In addition, in the past several years, 
our understanding for the systematic effects present in {\it Spitzer} time-series
observations has significantly increased. 
Motivated by the combination of these factors, we examine the available 
secondary eclipse spectroscopy data sets in a 
self-consistent and up-to-date manner in order to improve our understanding 
of the atmosphere of this hot Jupiter and test the \citet{gri08} results.

\capstartfalse
\begin{deluxetable}{ll}
\tabletypesize{\scriptsize}
\tablewidth{0pt}
\tablecaption{Adopted Stellar and Planetary Parameters for HD~189733\tablenotemark{a}}

\startdata
\hline\\ [-1.0ex]
M$_\star$ (M$_{\odot}$) & $0.823^{+0.022}_{-0.029}$ \\
R$_\star$ (R$_{\odot}$) & $0.766^{+0.007}_{-0.013}$\\
$\rm K_s$ (mag)\tablenotemark{b}&  $5.541\pm0.021$\\
T$_{\rm eff}$ (K)& $5090$\\
b$_{\rm impact}$ & $0.900^{+0.006}_{-0.010}$\\
M$_{\rm p}$ (M$_{\rm J}$)& $1.138^{+0.022}_{-0.025}$ \\
R$_{\rm p}$ (R$_{\rm J}$)& $1.178^{+0.016}_{-0.023}$ \\
P (days)              & $2.21857312^{+0.00000036}_{-0.00000076}$ \\
$\rm T_{0}$ (BJD$_{\rm TDB}$) & $2453988.804144^{+0.000072}_{-0.000039}$ \\
a$_{\rm p}$ (AU) & $0.03120^{+0.00027}_{-0.00037}$\\
e               & $0.0041^{+0.0025}_{-0.0020}$
\enddata
\tablenotetext{a}{Values from \citet{tri09}, except for $\rm K_s$ and T$_{\rm eff}$ \citep{knu10}.}
\tablenotetext{b}{Two Micron All Sky Survey (2MASS) $\rm K_s$ magnitude of the star (from the Infrared Science Archive: \\
\url{http://irsa.ipac.caltech.edu}).}
\label{tab:prop}
\end{deluxetable}
\capstarttrue

In Section~\ref{sec:obs} we discuss the available data. Section~\ref{sec:ana}
focuses on our analysis approach, while in Section~\ref{sec:dis} we discuss our 
results and compare them to theoretical models.

\section{Observations}
\label{sec:obs}
We analyze all 22 archival time-series spectroscopy data sets 
on HD~189733b observed with the {\it Spitzer}/IRS during secondary eclipse. 
These observations will remain unique until a new infrared telescope 
with spectroscopic capabilities in the mid-infrared becomes operational. 
The Spitzer Heritage Archive\footnote{\url{http://sha.ipac.caltech.edu/applications/Spitzer/SHA/}.} 
(SHA) lists all data sets, including proprietary ones, thus we are confident that 
our data includes all IRS spectroscopy on HD~189733b during secondary eclipse. 
Our data cover a wavelength range between 5 and 40\,$\mu$m at resolution 
$R \sim 100$. The information about the available data sets is summarized in \autoref{tab:specdata}. 
The principal investigator of all observations (program ID numbers 30473 and 40504) is Carl J. Grillmair. 
\capstartfalse
\begin{deluxetable}{llllll}
\tabletypesize{\scriptsize}
\tablewidth{0pt}
\tablecaption{HD~189733b Spectroscopic Observation Details}
\tablehead{
\colhead{Data} &
\colhead{AOR } &
\colhead{Observation} &
\colhead{Wavelength} &
\colhead{Exposure} &
\colhead{Spectra} \\
\colhead{set} &
\colhead{key\tablenotemark{a}} &
\colhead{date} &
\colhead{range ($\mu$m)} &
\colhead{time (sec)} &
\colhead{count} 
}
\startdata
g1  & 18245632 & 2006 Oct 21 & 7.4-14.0 & 14.7 & 900 \\
g2  & 20645376 & 2006 Nov 21 & 7.4-14.0 & 14.7 & 950 \\
g3  & 23437824 & 2008 May 24 & 7.4-14.0 & 61.0 & 280 \\
g4  & 23438080 & 2008 May 26 & 7.4-14.0 & 61.0 & 280\\
g5  & 23438336 & 2008 Jun 02 & 7.4-14.0 & 61.0 & 280\\
g6  & 23438592 & 2008 May 31 & 7.4-14.0 & 61.0 & 280\\
g7  & 23438848 & 2007 Oct 31 & 7.4-14.0 & 61.0 & 280\\
g8  & 23439104 & 2007 Nov 02 & 7.4-14.0 & 61.0 & 280\\
g9  & 23439360 & 2007 Jun 26 & 7.4-14.0 & 61.0 & 280\\
g10 & 23439616 & 2007 Jun 22 & 7.4-14.0 & 61.0 & 280\\
g11 & 23440384 & 2008 Jun 09 & 5.0-7.5  & 61.0 & 280\\
g12 & 23440640 & 2008 Jun 04 & 5.0-7.5  & 61.0 & 280\\
g13 & 23440896 & 2007 Dec 07 & 5.0-7.5  & 61.0 & 280\\
g14 & 23441152 & 2007 Nov 06 & 5.0-7.5  & 61.0 & 280\\
g15 & 23441408 & 2007 Nov 11 & 5.0-7.5  & 61.0 & 280\\
g16 & 23441664 & 2007 Nov 09 & 5.0-7.5  & 61.0 & 280\\
g17 & 23441920 & 2007 Nov 24 & 5.0-7.5  & 61.0 & 280\\
g18 & 23442176 & 2007 Nov 15 & 5.0-7.5  & 61.0 & 280\\
g19 & 23439872 & 2007 Nov 04 & 13.9-21.3& 122 & 140\\
g20 & 23440128 & 2007 Jun 17 & 13.9-21.3& 122 & 140\\
g21 & 23442432 & 2007 Dec 10 & 19.9-39.9& 122 & 140\\
g22 & 23442688 & 2007 Jun 20 & 19.9-39.9& 122 & 140\\
\enddata
\tablenotetext{a}{The Astronomical Observation Request (AOR) key that uniquely identifies the 
observation in the {\it Spitzer} Heritage Archive (\url{http://sha.ipac.caltech.edu/applications/Spitzer/SHA/}).}
\label{tab:specdata}
\end{deluxetable}
\capstarttrue

\section{Data Analysis}
\label{sec:ana}
\subsection{Spectroscopy Extraction}
\label{sec:specex}
We start our analysis with the Basic Calibrated Data (BCD) files produced by 
the {\it Spitzer} calibration software, version S18.18.0.
Following the discussion in \citet{eas10}, we convert the UTC-standard 
(which includes leap seconds as often as every six months) timing 
information included in the headers to the continuous 
Barycentric Dynamic Time (TDB) standard. Each IRS image 
contains $128\times128$ pixels and the spectra are dispersed in wavelength
approximately in the direction of the pixel columns. The spatial information is
along the pixel rows. For the $7.4 - 14\,\mu$m data, 
we elect to clip the first two and the last three pixel rows 
(rows 1-2 and 126-128, corresponding to the longest and shortest wavelengths, respectively), 
since they appear to be noisier and could
be subject to unknown systematic effects. Similarly, we clip the 
first two and the last seven rows for the $5.0 - 7.5\,\mu$m observations, which 
cover only rows between 1 and 80 out of 128 on the images.

For each data set, we correct for energetic particle hits by 
following the value of a given pixel as a 
function of time. We replace pixel values that are at least $4\sigma$ 
away from a running median of width 5 with that median. 
In this manner, we correct about 0.8\% 
($5.0 - 7.5\,\mu$m data), 0.4\% ($7.4 - 14\,\mu$m data)
and 0.7\% ($14 - 40\,\mu$m) of the pixels in every image. 

\begin{figure*}
\epsscale{0.8}
\plotone{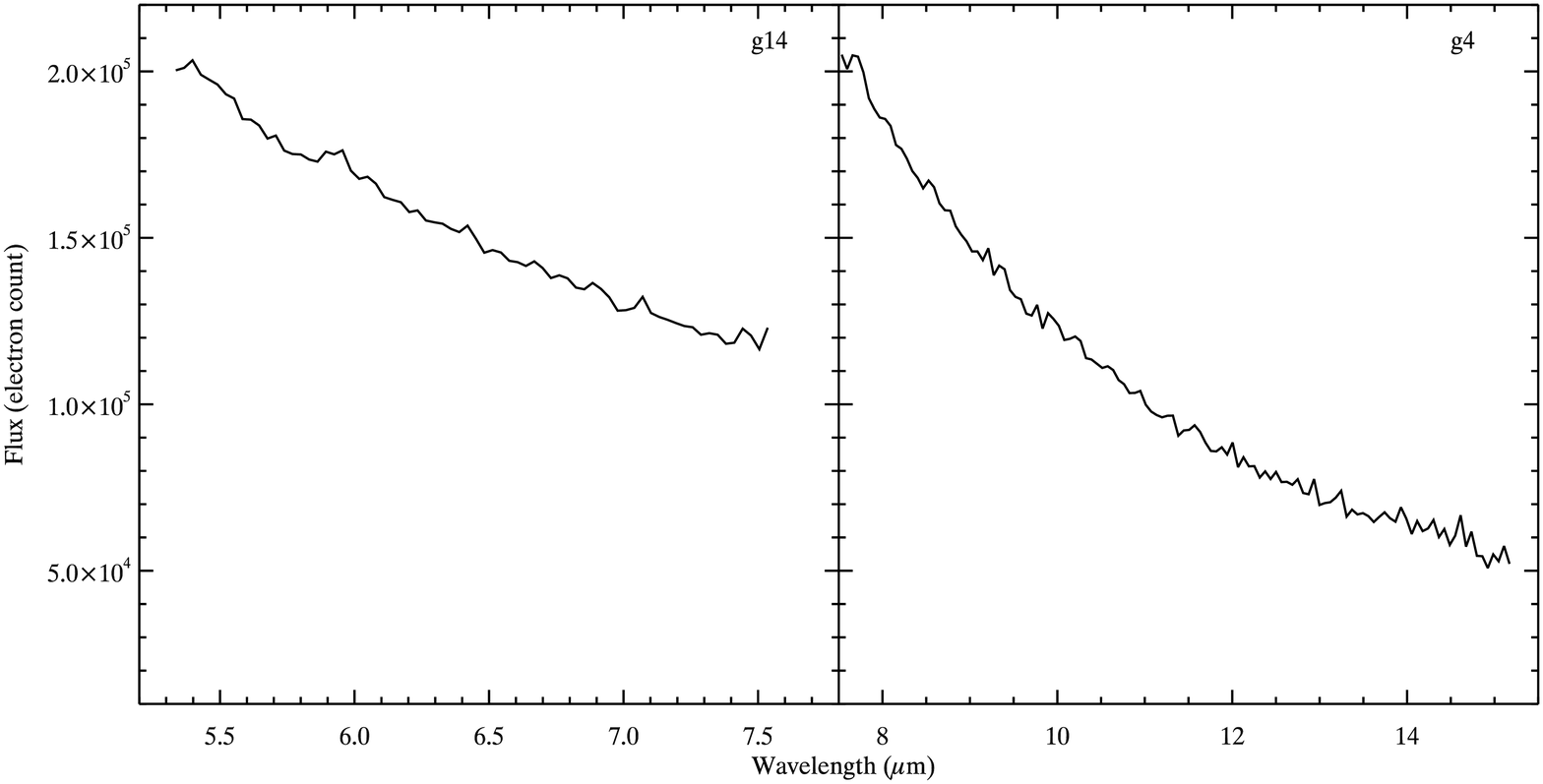}
\caption{
  Example single spectra derived using optimal extraction \citep{hor86}
  for two data sets -- g14 (left panel) and g4 (right panel). 
  The next steps of the analysis follow the detected flux at a given 
  wavelength as a function of time. 
}
\label{fig:raw_spec}
\end{figure*}

We employ optimal extraction \citep{hor86}, implemented in IDL, to reduce
the observed images to one-dimensional spectra. For each image, 
we estimate the background as a function of wavelength by fitting a 
third-order polynomial to every pixel row, i.e. along the 
spatial direction, excluding the region where the target was 
located. We determine the required polynomial order by experimenting with lower- and 
higher-order polynomials. Third order appears to be the lowest order polynomial 
that consistently fits the background. While second order polynomials often produce 
comparable fits, for some wavelengths they completely fail to match the background 
values in some region of 
the fitting domain. Thus, we elect third order polynomials to fit the background. 
Again, for each pixel row, we subtract the corresponding
background polynomial and then locate the peak of the source's brightness
by fitting a one-dimensional Gaussian. Thus, we are able to 
follow any gradual curvature in the shape of the dispersed spectrum 
on the image. We make an initial estimate of the detected brightness 
by integrating the background-subtracted flux values of each row. The integration 
is centered on the estimated brightness peak for that row and 
the integration range covers 3$\sigma$ in each direction around the
peak -- therefore changing with wavelength. We use this initial 
spectral estimate as input for the algorithm by \citet{hor86}, 
which iteratively calculates the optimum one-dimensional spectrum 
extracted from each image. As per the standard practice, the algorithm 
only uses the 3$\sigma$ span for the initial guess of the spectrum, 
but the iterations are performed over the whole spectral image. 
We show extracted sample spectra in \autoref{fig:raw_spec}.

\begin{figure*}
\epsscale{1.1}
\plotone{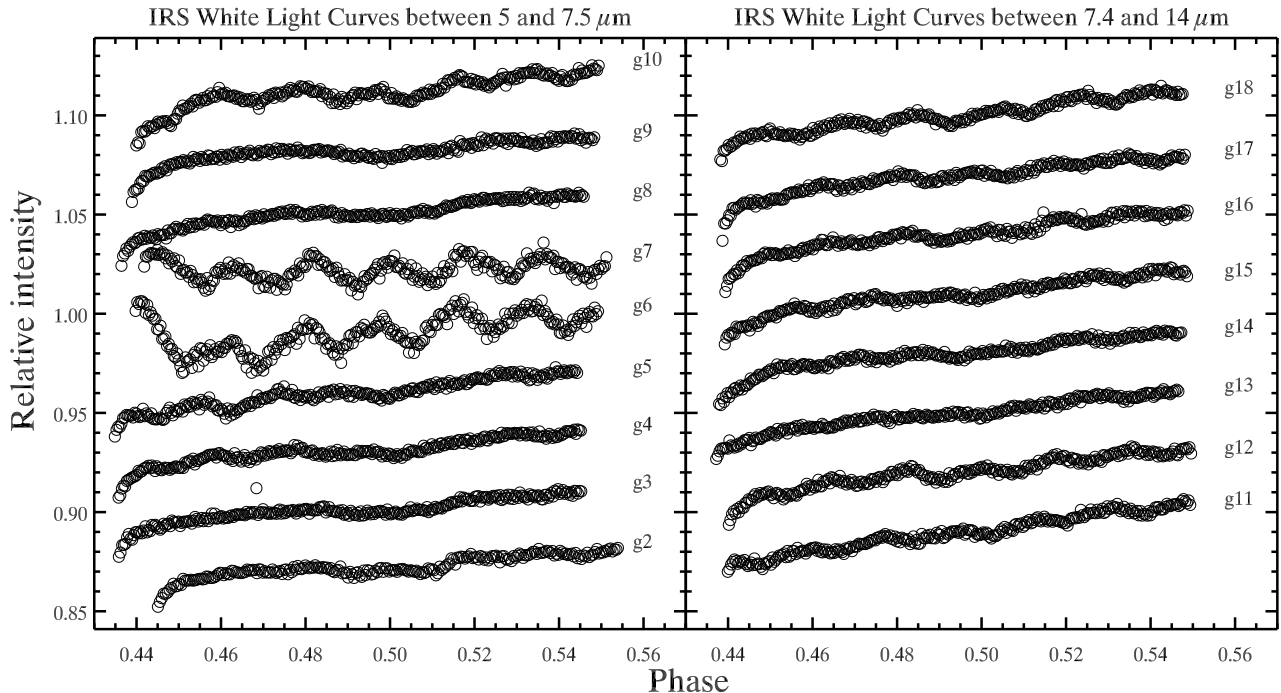}
\caption{
  The white light curves, shown here, for all data sets that
  we include in our analysis are calculated by integrating
  the extracted time series spectra from a given observation 
  over wavelength. The resulting white light intensity as
  a function of time is normalized to one at the time of 
  secondary eclipse and offset for clarity. We show both
  the 5-7.5$\mu$m (left) and 7.4-14$\mu$m (right) data. 
  The abscissa here represents time in units of orbital 
  phase. The timing information is extracted from the FITS
  file header as described in Section~\ref{sec:specex}. 
  We then convert the BJD$_{\rm TDB}$ times to phase by adopting 
  the ephemeris from \autoref{tab:prop}. The data sets are 
  labeled based on their designations adopted in 
  \autoref{tab:specdata}. One of the most obvious systematic 
  effects is the apparent intensity oscillation with time. This 
  is caused by a well documented telescope pointing jitter 
  \citep[e.g.,][]{gri07,dem11}. As the pointing shifts the IRS
  slit is illuminated by various portions of the target PSF, 
  resulting in changing apparent brightness. The systematic 
  effects present in the data are discussed in Section~\ref{sec:syseff}.
}
\label{fig:wlc}
\end{figure*}

Initially, we integrate each derived spectrum over wavelength and 
construct ``white light'' curves. We present these
for all data sets in \autoref{fig:wlc}, except for 
the g1 light curve (taken on 2006 Oct 21), which we exclude, because only for these 
data the telescope was nodded between two positions approximately 
every 12 minutes (30 exposures). The nod was performed in order to 
facilitate background removal, but this exacerbated the systematic
effects such as the ramp of intensity with time and the quasi-periodic
apparent flux variation (see Section~\ref{sec:syseff}). As a result, despite the 
identical exposure times, the g1 data set white light time-series 
exhibits higher noise than the g2 data set white 
light curve, where the telescope was not nodded.
We also exclude the g19 -- g22 data sets from the analysis (covering wavelengths
between 14 and 40$\,\mu$m), since these observations have a very 
low signal-to-noise ratio and do not even allow for useful eclipse 
depth upper limit determination. Thus, we focus on the g2 -- g18 data
sets. 

\subsection{Light Curve Analysis}
\label{sec:meas}
\subsubsection{Systematic Effects}
\label{sec:syseff}
We observe two major systematic effects that dominate the 
{\it Spitzer} IRS light curves. First, there is a positive
correlation of observed brightness with time, often called the 
ramp \citep{dem06}. This effect is similar to the ramp 
reported by {\it Spitzer} photometric time series studies 
at 8$\mu$m \citep[e.g.,][]{knu08,agol10,tod10}. It is thought
that the cause for this ramp might be that photo-electrons
are caught in quantum wells within the pixels and not 
read out at the start of the observation. As the 
observation progresses, most potential wells are filled
and most photo-electrons are correctly read, causing the
number of read-out electrons for a constant source to 
increase. Adopting this hypothesis, \citet{agol10} 
suggest a physically motivated toy model to account for 
the ramp, 
\begin{equation}
\label{eqn:ramp1}
\frac{F^\prime}{F} = a_0 - a_1e^{-t/\tau_1} - a_2e^{-t/\tau_2}. 
\end{equation}
Here $F^\prime$ is the observed intensity before the correction 
for the ramp, $F$ is the corrected intensity, $a_0$, $a_1$, $a_2$, $\tau_1$
and $\tau_2$ are free parameters and $t$ is time in arbitrary units. Here, 
we use time in units of orbital phase, where $t=0$ is the middle of 
primary transit, and $t=1$ occurs exactly one orbital period later. 
The coefficients in the first and second exponential terms are strongly correlated
and degenerate. Thus, including all parameters in the Markov
Chain Monte Carlo (MCMC) algorithm that we use to fit for the
eclipse depths causes convergence problems. Therefore, we elect
to neglect the second term in \autoref{eqn:ramp1}, 
\begin{equation}
\label{eqn:ramp1single}
\frac{F^\prime}{F} = a_0 - a_1e^{-t/\tau_1}. 
\end{equation}
This is justified since it results in only a marginal increase of 
the minimum $\chi^2$ values achieved by the fits, and a 
significant reduction of their Bayesian Information 
Criterion value \citep[BIC,][]{sch1978}.

We compare this to several other ramp correction functions that 
are not physically motivated but have been used in past studies, or are simple extensions of the single exponential function. 
We explore the log-linear \citep[e.g.,][]{tod10}, the log-quadratic \citep[e.g.,][]{cha08}, the third order polynomial
\citep{gri07}, which was used in the older analysis on part of our data, and the 
single exponent with a linear or quadratic functions added: 
\begin{eqnarray}
\label{eqn:ramp2}
\frac{F^\prime}{F} = a_0 - a_1t - a_2log(t),\label{eqn:linlog}\\
\frac{F^\prime}{F} = a_0 - a_1t - a_2t^2 - a_3log(t), \label{eqn:polylog}\\
\frac{F^\prime}{F} = a_0 - a_1t - a_2t^2 - a_3t^3,\label{eqn:3poly}\\
\frac{F^\prime}{F} = a_0 - a_1e^{-t/\tau_1} - a_2t, \label{eqn:lin1exp}\\
\frac{F^\prime}{F} = a_0 - a_1e^{-t/\tau_1}- a_2t - a_3t^2. \label{eqn:par1exp}
\end{eqnarray} 
The symbols for these expressions are defined as for 
\autoref{eqn:ramp1}. All of the functions tested
produce similar eclipse spectrum shapes, but the single exponential function typically produces 
the lowest BIC values in the individual wavelengths, especially at wavelengths
longer than $\sim8\,\mu$m. 
Thus, we elect to adopt it for the correction of the ramp in our 
analysis. While the choice of ramp correction 
function does not impact the overall shape of the planetary emission spectrum, it 
affects the absolute scale of the eclipse depths typically by $\sim20\%$. 
We discuss this effect further in Section~\ref{sec:old}.

The second systematic effect is a quasi-periodic
variation in detected flux with a period of about $60$\,min. 
This is caused by the well-known 
\citep[e.g.,][]{gri07,knu08,dem11} Spitzer pointing
oscillation. This has been traced to the periodic working cycle 
of a heater that keeps a battery within its operating 
temperature\footnote{\href{http://ssc.spitzer.caltech.edu/warmmission/news/21oct2010memo.pdf}{Spitzer Science Center
memo from 21 Oct 2010 (external link)}.}. 
In photometric data, the change in 
observed brightness of the target is caused by the variable 
sensitivity across the surface of a single pixel. For IRS {\it spectroscopy} data, 
however, the slit drifts with the pointing oscillation 
and samples parts of the stellar PSF that have different brightness. 
To put this in perspective, while the IRS 
slit is $3\farcs6$ wide, 
the angular resolution of the telescope varies between 
$1\farcs5$ and $4\farcs2$ in the $5-14\,\mu$m 
range\footnote{\href{http://irsa.ipac.caltech.edu/data/SPITZER/docs/irs/irsinstrumenthandbook/4/}{IRS Instrument Handbook (external link)}.}. 
This effect is easily seen in the
white light time series shown in \autoref{fig:wlc}. We 
refer to it as the ``sawtooth'' effect, following the shape
of the decorrelation function adopted by \citep{gri08} -- an 
asymmetric triangular sawtooth function with constant amplitude 
with time. However, in \autoref{fig:wlc} it is evident that
the amplitude of the sawtooth correction may be variable throughout
a single secondary eclipse observation. Thus, we explore alternative sawtooth removal
methods. 

Ideally, if precise pointing information was known for the 
telescope during the time of the observations, the pointing
oscillation, and thus the shape of the sawtooth correction 
function, could be reconstructed from first principles. However, 
the pointing information recorded in the FITS file headers show 
no pointing jitter, nor does the raw spacecraft telemetry 
that was graciously provided to us by the {\it Spitzer} Science Center HelpDesk and Sean Carey. 
It is possible that the temperature changes due to the operation 
cycle of the heater responsible for this effect cause the 
observatory star trackers to begin to lose and then regain their alignment
with the telescope's boresight. Thus, despite the real 
pointing oscillation, the star trackers record no shift in the
direction at which the main mirror is pointed. For the same reasons, 
the spacecraft gyroscopes also do not provide any useful information.

Another option is to centroid on a star whose image falls on 
one of the IRS peak-up arrays and is imaged during every IRS 
exposure. Unfortunately, there are no fortuitous high 
signal-to-noise observations of point sources recorded in the IRS peak-up 
images. 

\begin{figure}
  \epsscale{1.1}
  \plotone{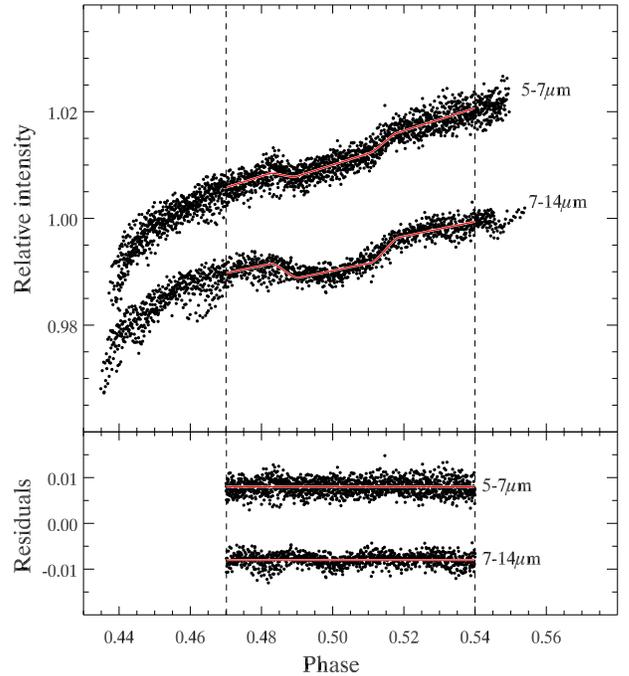}
  \caption{
    Upper panel: Stacked white light curves in the 5-7 and 7-14$\,\mu$m ranges, 
    normalized to unity and arbitrarily offset for clarity. In the combined white light 
    curves, the sawtooth effect is largely canceled. The red lines indicate
    the best fit results from our Markov Chain Monte Carlo (MCMC) analysis, including 
    a ramp with time and a secondary eclipse, but not the sawtooth. The phase coverage
    of the red line and the dashed lines denote the data points that were included in the MCMC fits. 
    Lower panel: The residuals achieved by subtracting the best fit eclipse and ramp models from the 
    stacked white light curves, again arbitrarily offset for clarity. 
    The red lines indicate the zero levels for the corresponding white light residuals.
  }
  \label{fig:wlc_stack}
\end{figure}

However, there are multiple spectroscopic light curve observations 
of HD~189733 for each wavelength range. Since the amplitude and
phase of the sawtooth variation are unrelated to the astronomical 
observations, they are different for the different light curves. 
Hence, by stacking the white light curves together, we are able 
to nearly average out the sawtooth variations. When combining, we omit the g6, g7
and g10 data sets due to their relatively high sawtooth amplitudes. 
This results in high signal-to-noise light curves, in which the dominant 
systematic effect is the ubiquitous Spitzer ``ramp'' and the sawtooth
contribution is minimized. We show these curves in \autoref{fig:wlc_stack}.

Measuring the broadband eclipses and the ramps in the white light {\it stacked} light 
curves allows us to subtract them from the {\it individual} white light curve observations
(shown in \autoref{fig:wlc}), which leaves us with 17 residual curves, 
corresponding to each eclipse observation. We smooth these to extract the shape 
of the sawtooth correction that we need for each
data set. The pointing oscillation of the telescope does not depend on wavelength, but the 
size of the stellar PSF compared to the size of the slit does. Therefore, 
the shape of the sawtooth correction is independent of wavelength, 
but the amplitude of the sawtooth {\it does} vary with wavelength. 
Thus, if we multiply the derived sawtooth correction 
function by a wavelength-dependent scaling factor, we can apply it to 
individual single-wavelength light curves. An advantage of this method is
that it makes no assumptions about the shape of the sawtooth. However, 
we lose any information about the variability of the
planetary emission, since we stack the white light curves, implicitly 
assuming that the broadband eclipse depths are equal in all light curves. 
Since \citet{agol10} place a $1\sigma$ upper limit on HD~189733b's 
variability at 8$\,\mu$m of 2.7\%, any variability is likely to be 
below the {\it Spitzer}/IRS detection limit. Thus, the constant flux 
assumption is justified for this planet. 

The lower panel of \autoref{fig:wlc_stack} still shows some residual red noise, 
especially in the 7-14$\,\mu$m range. This could bias the fit towards 
a slightly deeper or shallower eclipse measurements than the ``true'' value. 
Since we are subtracting the best fit eclipses from the stacked white light curves
from the individual light curves, this implies including a small constant positive
or negative eclipse in the sawtooth correction, leading to a small offset in the
measured secondary eclipses (i.e., planetary intensity) in individual wavelengths. 
We stress that this effect, if it occurs, should have no impact on the 
shape of the planetary spectrum. 
We describe the details of our sawtooth correction algorithm in detail 
in the next section. 

\subsubsection{Fitting Procedure}
\label{sec:fp}
In order to stack the light curves correctly, 
we convert the observation time for each spectro-photometric data 
point in all data sets to units of orbital phase using the ephemeris given in 
\autoref{tab:prop} \citep{tri09}. We then normalize
each light curve so that its average flux is unity during the expected time of 
secondary eclipse. Next, we simply combine all white light photometric
measurements in a single data set, where the data points are ordered by their 
calculated orbital phase. Here, we bin the points from the g2 data 
by four points, because the exposure times for this
observation were about four times shorter than for the other data sets
(see \autoref{tab:specdata}). For the stacked white light curves, we 
adopt the following expression as a model to fit to the data: 
\begin{equation}
\label{eqn:wlc}
I(t) = a_0 - a_1e^{-t/\tau_1} + d_1M_{c}.
\end{equation} 
Here, $d_1$ is the eclipse depth, $a_0$, $a_1$ and $\tau_1$ are the
ramp parameters from \autoref{eqn:ramp1single}. The eclipse shape, $M_{c}$, 
centered on a given central phase, $c$,
is based on the \citet{man02} model but without limb darkening (since 
the planet is behind the star, the bottom of a secondary eclipse light 
curve is flat). The duration of the eclipse and ingress/egress is fixed, 
based on the planetary and stellar radii, eccentricity and orbital period 
adopted in \autoref{tab:prop}. We set $a_0$, $a_1$, $\tau_1$, $d_1$ and $c$ to be 
the free parameters for our stacked white light curves fit. 

Not all data sets were observed starting at the exact same orbital phase, and 
hence the steep part of the ramp occurs at slightly different phase for each 
light curve. This makes stacked light curves noisy at phases earlier than 
$\sim0.47$. In addition, not all data sets cover orbital phases higher 
than $\sim0.54$, causing the stacked white light curves to be vulnerable 
to poor sawtooth cancellation for phases above this threshold. Thus, we
only use data with phases between $0.47$ and $0.54$  for the fit of the 
stacked white light curves, and the stacked light curves at individual 
wavelengths described in forthcoming sections.  

We implement a MCMC fitting routine following
the algorithm outlined by \citet{ford05,ford06}. We perturb only one, 
randomly selected, free parameter at a time. We perform $6\times10^6$ 
MCMC steps ($\sim10^6$ steps per parameter). Of the total length of the chain, 
we drop the initial $10^6$ steps as
``burn-in'' time, required for the chain to converge. 
Before running the long chain, we run several shorter chains 
in order to optimize the widths of Gaussian 
distributions that are used to determine the size of the parameter 
perturbation. We elect these widths to be such that the acceptance rate
of the new parameter value to be between 35\% and 55\% in order to 
optimize the efficiency of convergence \citep{ford06}. The histograms
of the parameter runs closely resemble Gaussians, therefore we adopt the 
mean values for the eclipse depth and the eclipse central phase
as the best fit values for these MCMC fits. The MCMC states that
result in the minimum $\chi^2$ values have eclipse depths typically 
within $3\%$, or much less, of the mean histogram value. Therefore, 
the choice of ``best value'' does not change our final results. The best fit eclipse
depth in the 5 -- $7\,\mu$m range data is $0.216$\% with a central 
phase of $0.50014$. For the 7 -- $14\,\mu$m-range stacked white 
light curve, the eclipse depth is $0.370$\% and the central phase is
$0.50060$. Just due to the light travel time delay of the HD~189733b 
system, and adopting $e\cos\omega = 0$ \citep{agol10}, the expected
central phase is $0.50016$. In addition, \citet{agol10} detect an 
additional $38 \pm 11$\,s ($0.00020 \pm 0.00005$ in units of phase)
delay that they attribute to the hottest point of the day side of the
planet lagging behind along the orbit compared to the substellar point. 
Hence, our white light central phase measurements are compatible with 
previous studies, after allowing for the fact that at different 
wavelengths the delay due to hot spot offset may be different, since
we are probing different atmospheric altitudes. 

\begin{figure}
  \epsscale{1.1}
  \plotone{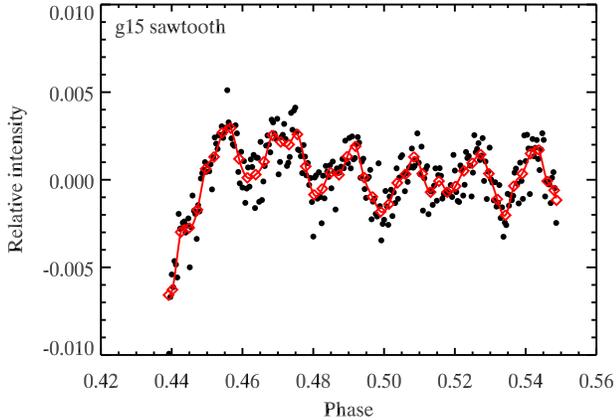}
  \caption{
    The residuals after subtracting the ramp and eclipse from the 
    g15 white light curve (black points). Since this data set covers
    the 5 -- 7$\,\mu$m range, we fixed the eclipse depth at 0.216\%
    and the central phase at 0.50014. The residuals were binned by 
    six (red diamonds), and a cubic spline interpolation was 
    used to estimate the final sawtooth function at the observed
    orbital phases (solid red line).  
  }
  \label{fig:saw}
\end{figure}

In order to estimate the sawtooth correction function for each 
individual data set, we use the same MCMC algorithm as above to 
fit a ramp (but still no sawtooth) to the individual white light 
curves shown in \autoref{fig:wlc}. This time, we hold the 
eclipse depth and central phase fixed to the values derived from 
the corresponding stacked white light curves. As above, we use 
the ramp parameters from \autoref{eqn:ramp1single}. We subtract 
the best fit ``ramp-and-eclipse'' model from the white light curve
and we are left with residuals that represent the sawtooth function 
combined with the photon noise. In order to eliminate the photon 
noise and estimate the sawtooth, we bin the residuals by six. We
choose this factor empirically -- bigger bins smooth the 
sawtooth curve too much and degrade the fits, while smaller bins are 
dominated by the photon noise. We then utilize the cubic spline 
interpolation IDL routine {\it spline} to evaluate the sawtooth function at the orbital 
phases when the data were actually observed. A sample sawtooth 
function is presented in \autoref{fig:saw}.

The light curves derived from a single pixel row (the 
wavelength dispersion direction is along the image columns)
have a very low signal-to-noise ratio. To improve this, we bin
the light curves in bins of width three pixels in {\it wavelength}, such that
a given wavelength channel is the combination of three 
image rows, not just one. In this way, we are left with 24 channels
between $5\,\mu$m and $7\,\mu$m, and 42 channels between $7\,\mu$m 
and $14\,\mu$m, per data set. We stack all light curves in 
a given wavelength channel in the same way we stack the
white light curves (but this time including the g6, g7 and g10 data 
sets). Again, the g2 data points are binned by a factor of four
to account for their shorter exposure times. Since we have estimated
the value of the sawtooth function for each data set and 
for each orbital phase, we stack the sawtooths in the same way 
we stack the light curves -- by simply combining the data in 
a single data set and ordering the points in it by their orbital 
phase. In this way each photometric point in the stacked light curves
will be corrected by the corresponding sawtooth value of its original 
data set. 

The amplitude of the sawtooth correction is dependent on wavelength, as
discussed in Section~\ref{sec:syseff}. To account for this, we introduce
a new free parameter, the sawtooth scale, $\xi$. We assume that $\xi$
depends only on wavelength, and that there is a single value of 
$\xi$ that applies to all data sets at a given wavelength channel, i.e., that it 
is independent of time. Because the non-stacked 
light curves for any given channel are still relatively noisy, this assumption is difficult to 
test in practice. However, it is reasonable, since the width of the 
target PSF compared to the slit width is what determines the sawtooth
amplitude resulting from a given pointing shift, and this is independent 
of time, and only depends on the wavelength. Variable PSF width can 
change this, but {\it Spitzer} is
on an Earth-trailing orbit, and is far more thermally stable than, e.g. 
the {\it Hubble Space Telescope} on its low-Earth orbit. Therefore, it is
not surprising that we see no significant changes in the point response
function (PRF) of the telescope that might indicate a change in the
relative sizes of the slit and the PSF at a given wavelength. 

\subsubsection{Best Fit Depths}
\begin{figure}
  \epsscale{1.1}
  \plotone{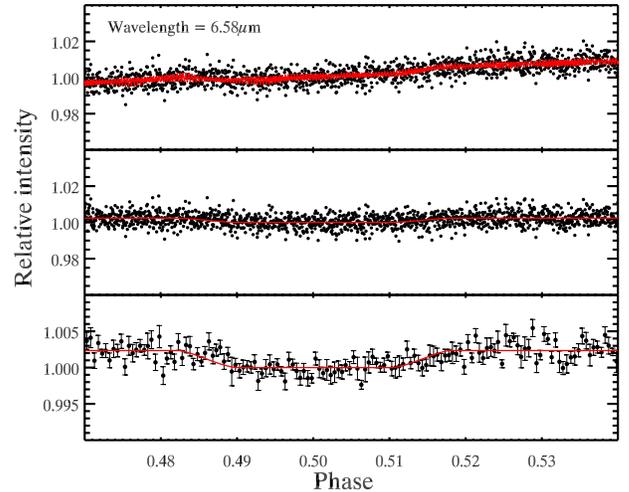}
  \caption{
    The raw (top panel, black points) stacked light curves at $6.58\,\mu$m based 
    on all nine available data sets that contain this wavelength channel are fitted 
    with a model (red line) including the eclipse depth, the {\it Spitzer} ramp and 
    the sawtooth correction. In the middle panel, the light curve (black points) has been 
    corrected for the systematic effects, with the best fit eclipse model shown with the 
    red line. The bottom panel is similar, but here the corrected light curve is binned, each bin
    with length of $0.0005$ in units of phase ($1.6\,$min, or $\sim10$ data points). 
    The vertical axis is rescaled to emphasize the eclipse, which is 
    clearly visible in the binned light curve. We exclude the intensity measurements with
    orbital phases below $0.47$ and above $0.54$, as discussed in Section~\ref{sec:fp}.
  }
  \label{fig:wv20}
\end{figure}

\begin{figure}
  \epsscale{1.1}
  \plotone{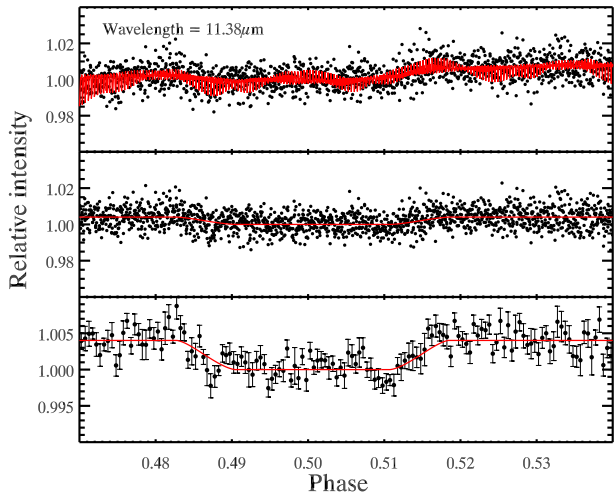}
  \caption{
    Similar to \autoref{fig:wv20}, but here we show the $11.38\,\mu$m 
    light curve, composed using all eight data sets that include this wavelength. 
    The bins in the bottom panel, again, cover $0.0005$ in units of phase, or 
    $1.6\,$min, but for these data this corresponds to $\sim11$ points per bin. 
  }
  \label{fig:wv10}
\end{figure}

We perform both MCMC and Prayer Bead Monte Carlo 
\citep[PBMC,][]{gil07} fits for the stacked
light curves at each of the 56 wavelength channels. Our MCMC fits utilize the same
algorithm described in Section~\ref{sec:fp}, except here we include 
the sawtooth correction. We model the light curves with the 
following expression, 
\begin{equation}
\label{eqn:mcmc}
\frac{F^\prime}{F} = a_0 - a_1e^{-t/\tau_1} + {\xi}S_s + d_1M_c. 
\end{equation}
$S_s$ represents the stacked sawtooth correction function. The free 
parameters in this fit are $a_0$ ,$a_1$ , $\tau_1$ , $\xi$ and $d_1$. 
We fix the central phase to the best fit value from the stacked white light
curve fits since it is approximately independent of wavelength\footnote{
The hot spot offset from the substellar point inferred by \citet{knu07} 
and \citet{agol10} at 8$\,\mu$m, to which the eclipse central phase delay 
is attributed to, may be different at different atmospheric altitudes. 
This, combined with the fact that different observing wavelengths probe
different layers of the atmosphere, means that the apparent delay of 
the secondary eclipse central phase may be dependent on wavelength. However, 
this effect is most likely very subtle and can safely be ignored here.}.
For each wavelength channel, we record the MCMC free parameter runs and 
create a histogram of the values for each parameter that closely resemble
Gaussian distributions. We adopt the mean of the histogram to be 
the ``best fit''. 
As before, using the minimum $\chi^2$ eclipse depths does not change
our ``best fit'' values by more than $\sim3\%$, much less than our uncertainties 
and the choice of best fit value has no impact on the final results.
We show typical raw and corrected light curves for two wavelength 
channels, along with the best fit MCMC models for the eclipse and the
systematic noise in Figures~\autoref{fig:wv20}~and~\autoref{fig:wv10}.

It is possible that our sawtooth correction method does not remove the 
sawtooth perfectly, and there could be additional, less important, 
systematic effects that we have not taken into account. Therefore, there
may be residual red noise in the systematics-corrected light curves (e.g.,
middle and lower panels of Figures~\autoref{fig:wv20}~and~\autoref{fig:wv10}).
Similarly to the photometric analyses in, e.g., \citet{des11,dem11,tod12,tod13}, in order to 
account for the possible influence of the residual correlated noise, we
perform PBMC fits for the light curves observed at each wavelength channel. 
The PBMC is based on simulating additional data sets by 
subtracting the best fit residuals from the observed data. The 
residuals are typically shifted over by a fixed number and added
back to the best fit. This creates a simulated observation where
the red noise from the original data is preserved. Simulating multiple 
data sets in this manner and fitting the eclipse depth model to them 
allows us to quantify the impact of correlated noise on our results, 
unlike the MCMC fitting routine, which assumes Gaussian uncertainties. 

For the PBMC fits, we adopt the MCMC model realization 
with the smallest $\chi^2$-statistic to be the best fit model.  
Shifting the residuals of the stacked light curves with respect to the
best fit model is inappropriate in this case, because each observed eclipse
within the stacked light curve may have its own associated red noise, 
independent of that of the other observations. Therefore, instead of
shifting the stacked light curve residuals by one at every PBMC iteration, 
we calculate the residuals of each individual light curve at
a given wavelength separately and shift each of them by a random number. Then we add the
shifted residuals back to the best fit model and stack the thus 
simulated light curves the same way as the original data. Performing an MCMC fit to the simulated
stacked light curves is extremely computationally expensive. Therefore, 
we fit the model using the IDL {\it mpfit} package for non-linear
least squares fitting \citep{mar09}. The free parameters used here 
are the same as the ones used for the observed data MCMC fit. In this way, 
we simulate 10,000 stacked light curves per wavelength channel, and 
this allows us to evaluate the cumulative effect that the residual 
correlated noise may have on the eclipse depth results. We create histograms 
from the eclipse depths from the PBMC runs for each wavelength. 
These are typically close to Gaussian in shape, but often with some asymmetry 
as expected for data where small amounts of red noise remain uncorrected. 
Since the astrophysical 
white noise is unrelated to the telescope systematic noise, any of the simulated 
PBMC realizations could have been the {\it observed} data. Therefore, we choose 
to adopt the medians of these histograms instead of the MCMC best fit values 
as the ``best fit'' points in the histograms. 

\subsubsection{Uncertainty Estimates}
We fit each MCMC histogram with a Gaussian function
and adopt its standard deviations to be the uncertainty of 
the parameter at the given channel.
We define the $1\sigma$ PBMC uncertainties as the region that covers 68\% of the histogram 
centered on the median simulated-data eclipse depth. We select this definition 
because it allows for efficient comparison on an even footing with the results 
from the MCMC Gaussian uncertainties. 
We compare the uncertainties from the PBMC and MCMC estimates of the
eclipse depths in all wavelength channels. We find that the PBMC 
uncertainties are, in all cases, larger, but the PBMC ``best values''
(the histogram medians) are always well within $1\sigma$ of the 
MCMC best fits. The MCMC uncertainties are smaller than the 
PBMC ones by between $14$ and $69\%$, but typically 
by about $50\%$. 

Both the MCMC and PBMC errors give good estimates of how the magnitude of 
the error bars varies with wavelength, due to changing photon counts per exposure per wavelength
 -- their wavelength dependence is very similar.
However, they are imprecise estimates of the errors in an absolute sense -- 
the MCMC uncertainties do not take residual red noise into account, while the 
PBMC uncertainties give large uncertainties to account for the uncertainty in the 
absolute eclipse depths, even though they do not account for systematic effects, like the 
choice of ramp correction function (Sections~\ref{sec:syseff}~and~\autoref{sec:old}). 
Thus,the PBMC uncertainties overestimate
the uncertainty in the difference between two wavelengths. 
Since the overall absolute contrast may vary depending on the choice of a ramp function, 
we focus on the point-to-point changes in the spectrum. 
Therefore, we use the PBMC and MCMC uncertainties to 
describe the wavelength variation of the errors, but we exploit the nature of the spectrum to 
scale all of the the errors accurately to reflect their correct wavelength-to-wavelength uncertainty. 

For any measurement, the best way to 
estimate the real magnitude of the errors is to repeat the measurement many times independently 
and compare the different results. This is impractical to do for eclipse photometry
\citep[except in a few cases, e.g.,][]{agol10}, but the nature of low- to moderate resolution spectra 
can be exploited as a proxy for independent measurements. Because the spectrum changes 
only slowly with wavelength compared to the precision of our data,  the eclipse depth should 
be closely the same at adjacent wavelengths. Hence, we use the wavelength to wavelength 
differences in derived eclipse depth to scale the errors in an absolute sense.

To scale the uncertainties, we adopt the following procedure: first, we take the 
HD~189733 star-planet contrast model from Burrows that we present in Section~\ref{sec:bur} and 
we bin it to the resolution of our extracted IRS spectrum. We find that the binned model,
varies smoothly from spectroscopic channel to channel. We express the wavelength dependence 
of the uncertainties (longer wavelengths yield higher uncertainties)
by taking the average of the PBMC and the MCMC uncertainties 
for a given wavelength, $\zeta_i$. We add random Gaussian noise to the binned Burrows model, 
with $1\sigma = k\zeta_i$ at the respective wavelengths, where $k$ is a
scaling factor. In this way, we simulate observations for a range of values for $k$, 
varying it in steps of 0.01. We estimate the point-to-point scatter of the observed and simulated 
eclipse spectra by taking the standard deviations of their derivatives. We
find that they are closest for $k=0.887$. Thus, as final uncertainties, we adopt 
the average of the PBMC and the MCMC uncertainties, $\zeta_i$, scaled by $0.887$. 
We stress that these values refer to the 
point-to-point uncertainties only, which determine the shape of the spectrum, not
to the absolute eclipse depths, which can be affected by additional 
systematic effects, as discussed below.  

We examine the possibility for placing an upper limit on the time-variability of the planet's 
emission in different wavelengths. However, the uncertainties of the individual 
eclipse observations as a function of time are too large to place meaningful constraints
on this quantity in a way similar to the 8$\,\mu$m multi-epoch photometric study by \citet{agol10}.

Another interesting caveat is that in reality the spectral coverage of the
$7-14\,\mu$m IRS spectra extends slightly beyond $15\,\mu$m. The
intensities observed at wavelengths above $\sim13.5\,\mu$m are affected
by the well-documented ``teardrop'' 
effect\footnote{\href{http://irsa.ipac.caltech.edu/data/SPITZER/docs/irs/features/\#8_SL1_14u_Teardrop}
{IRS Features and Caveats: SL 14$\,\mu$m Teardrop (external link)}.}, 
which causes the spectral trail to appear ``lumpy'' on the images, like a teardrop. The cause of this systematic
effect is either light leakage, or defects in the detector or optics. 
There is no reliable correction, and therefore we exclude all eclipse depths at wavelength longer than 
$\sim13.5\,\mu$m from our results. We summarize the 56 final eclipse 
depths and their uncertainties in \autoref{tab:res}. 

\capstartfalse
\begin{deluxetable}{cc}
\tabletypesize{\scriptsize}
\tablewidth{0pt}
\tablecaption{Secondary Eclipse Spectroscopy Results}
\tablehead{
\colhead{Wavelength}&
\colhead{Eclipse}  \\
\colhead{($\mu$m)}&
\colhead{Depth (\%)}
}
\startdata
IRS $5-7\,\mu$m mode: & \\
$5.46$ & $0.211\pm0.031$\\
$5.55$ & $0.195\pm0.036$\\
$5.65$ & $0.195\pm0.026$\\
$5.74$ & $0.222\pm0.023$\\
$5.83$ & $0.251\pm0.037$\\
$5.92$ & $0.190\pm0.029$\\
$6.02$ & $0.198\pm0.023$\\
$6.11$ & $0.214\pm0.028$\\
$6.20$ & $0.256\pm0.039$\\
$6.30$ & $0.243\pm0.027$\\
$6.39$ & $0.227\pm0.026$\\
$6.48$ & $0.252\pm0.023$\\
$6.58$ & $0.192\pm0.029$\\
$6.67$ & $0.194\pm0.027$\\
$6.76$ & $0.194\pm0.028$\\
$6.85$ & $0.239\pm0.030$\\
$6.95$ & $0.189\pm0.032$\\
$7.04$ & $0.240\pm0.031$\\
$7.13$ & $0.226\pm0.029$\\
$7.23$ & $0.292\pm0.038$\\
$7.32$ & $0.207\pm0.041$\\
$7.41$ & $0.318\pm0.049$\\
$7.51$ & $0.227\pm0.068$\\
\hline\\ [-1.5ex]
IRS $7-14\,\mu$m mode: & \\
$7.53$ & $0.314\pm0.050$\\
$7.66$ & $0.287\pm0.036$\\
$7.84$ & $0.257\pm0.034$\\
$8.03$ & $0.296\pm0.034$\\
$8.22$ & $0.266\pm0.028$\\
$8.40$ & $0.324\pm0.034$\\
$8.59$ & $0.317\pm0.032$\\
$8.77$ & $0.343\pm0.034$\\
$8.96$ & $0.318\pm0.036$\\
$9.15$ & $0.366\pm0.037$\\
$9.33$ & $0.314\pm0.030$\\
$9.52$ & $0.333\pm0.035$\\
$9.71$ & $0.382\pm0.038$\\
$9.89$ & $0.362\pm0.042$\\
$10.08$ & $0.377\pm0.035$\\
$10.27$ & $0.420\pm0.036$\\
$10.45$ & $0.340\pm0.038$\\
$10.64$ & $0.331\pm0.050$\\
$10.82$ & $0.418\pm0.045$\\
$11.01$ & $0.394\pm0.038$\\
$11.20$ & $0.417\pm0.041$\\
$11.38$ & $0.401\pm0.048$\\
$11.57$ & $0.357\pm0.035$\\
$11.76$ & $0.408\pm0.046$\\
$11.94$ & $0.416\pm0.041$\\
$12.13$ & $0.367\pm0.044$\\
$12.31$ & $0.522\pm0.040$\\
$12.50$ & $0.473\pm0.047$\\
$12.69$ & $0.443\pm0.042$\\
$12.87$ & $0.383\pm0.047$\\
$13.06$ & $0.378\pm0.045$\\
$13.24$ & $0.461\pm0.075$\\
$13.43$ & $0.365\pm0.046$
\enddata
\label{tab:res}
\end{deluxetable}
\capstarttrue

\section{Discussion}
\label{sec:dis}
\subsection{Comparison to Previous Studies}
\label{sec:old}
\begin{figure}
  \epsscale{1.0}
  \plotone{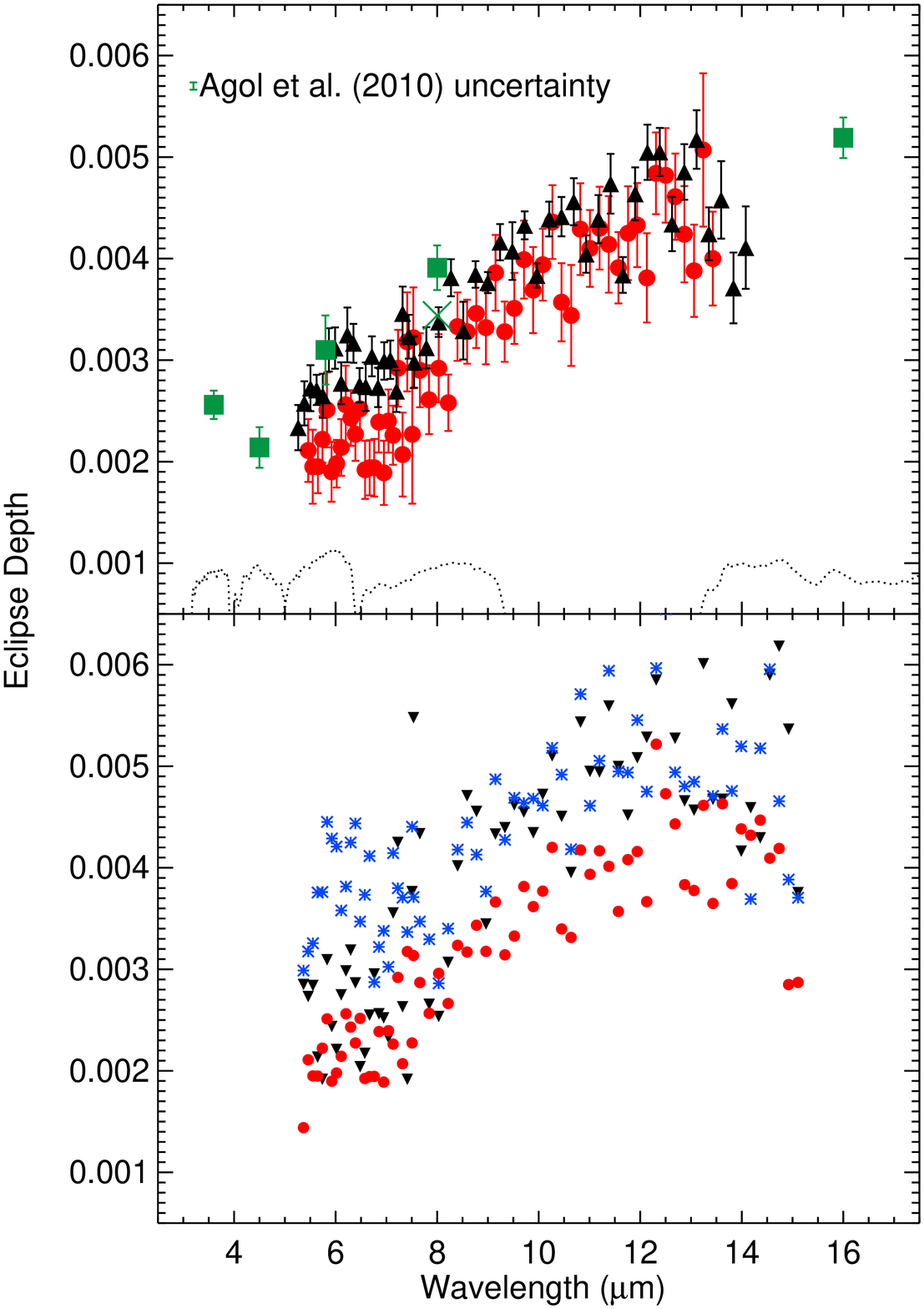}
  \caption{Top panel: We present a comparison between the eclipse depth spectrum derived here 
    (filled red circles) with the results from \citet{gri08} (black triangles) and the
    \citet{cha08} and \citet{agol10} secondary eclipse photometry (green squares and
    green $\times$-symbol, respectively). The size of the uncertainty in the \citet{agol10}
    measurement is indicated in the upper left corner. Bottom panel: comparison between our adopted 
    results (filled red circles) and analyses of our data using: a third order polynomial 
    ramp correction function \citep[black upside down triangles,][]{gri07} and the \citet{cha08}
    logarithm plus quadratic ramp correction function (blue asterisks), discussed in Section~\ref{sec:syseff}.
    The results based on the ramp functions in \autoref{eqn:linlog}, \autoref{eqn:lin1exp} 
    and \autoref{eqn:par1exp} show similar offsets from the
    results we adopt, and are not shown here for clarity. 
    The choice of ramp function affects the absolute eclipse depths, but not their overall shapes. Consistently
    with the previous studies, both of the ramp functions used in the two previous 
    studies yield deeper eclipses than the exponential ramp adopted here. 
  }
  \label{fig:comp_old}
\end{figure}

The eclipse depths we measure via secondary eclipse spectroscopy appear
systematically below the broad band photometry measured by \citet{cha08} 
and \citet{agol10}. \citet{gri08} also find deeper eclipses than this study 
(by $\sim20\%$), especially below wavelengths of $\sim7.5\,\mu$m. As seen in 
the top panel of \autoref{fig:comp_old}, the difference decreases and becomes marginal for 
wavelengths above $\sim7.5\,\mu$m.

There are several possible explanations for this discrepancy. For instance, 
the host star HD~189733 has some variability. Then, it is possible that the observations presented by 
\citet{cha08} and \citet{gri08} were mostly performed near stellar flux minima, leading to 
deeper apparent eclipses, while the ones analyzed here in addition could have been observed 
near stellar flux maxima causing the eclipses to appear shallower. The \citet{agol10}
point is based on six separate eclipse depth measurements at $8\,\mu$m, and lies closer
to our results than the \citet{cha08} result at this wavelength. The variability of
HD~189733 is caused by large stellar spots that cover $\sim 1-2\%$ of the stellar surface \citep{hen08}
and are $\sim1000\,K$ cooler than the rest of the photosphere \citep{pon07}. They lead to a flux
variability of around $\sim 1-2\%$ in the visible. Assuming that the spots and the photosphere
have blackbody fluxes, a stellar variability of $1.5\%$ at $0.5\,\mu$m corresponds to 
variability of about $0.4$\% to $0.6$\% between $3.6$ and $16\,\mu$m. This is insufficient to 
explain the discrepancy between the results presented here and the \citet{cha08} and 
\citet{gri08} findings.

Another possibility is that the discrepancy is due to a systematic difference between
the sawtooth correction algorithms used in \citet{gri08} and in this study. Any bias or offset in the
sawtooth determination can lead to removing or adding to the eclipse depths. This is a more 
significant problem at the shorter wavelengths, between $5$ and $\sim7.5\,\mu$m, where the sawtooth 
amplitude is higher, perhaps explaining why the difference becomes small at the long wavelengths.
However, this effect is likely to be relatively small. 

The most likely reason for this is the choice of a ramp correction function (Section~\ref{sec:syseff}). 
The way the ramp correction function bends at the time of eclipse can be critical for the
absolute scale of the eclipse depth, as seen in the bottom panel of \autoref{fig:comp_old}.
While the shape of the spectrum is unchanged regardless of the choice of ramp correction function, 
substituting the physically motivated ramp 
correction function we adopt in our analysis with the functions used by \citet{gri07} and \citet{cha08}, a third
order polynomial (\autoref{eqn:3poly}) and a logarithm plus quadratic (\autoref{eqn:polylog}), respectively, yields results comparable with
the \citet{gri08} absolute depths. We suggest that a consistent re-analysis of the {\it Spitzer}/IRAC 
photometry utilizing the single exponent ramp correction function could yield somewhat lower
eclipse depths and could lead to a better agreement between the photometric points and 
the updated spectroscopy presented here. However, this analysis lies outside the scope of this work. 

The offset in absolute eclipse depth, likely caused by the choice of ramp correction, 
is not critical for our results because it appears to 
change slowly with wavelength, thus generally preserving the shape of the spectrum. It does, 
however, limit our ability to draw conclusions about the day-to-night side energy transfer 
efficiency, since this is the property that determines the overall dayside planetary flux levels. 
However, an examination of the shape of the P-T atmospheric profile is still possible. 

\subsection{Comparison to Emergent Spectrum Models}
\label{sec:mods}
In order to study the atmospheric properties of HD~189733b, we compare the measured emergent spectrum
to models. We utilize a simple and computationally efficient radiative transfer code
developed by \citet{ric03}, which is sufficient to retrieve the basic thermal properties of 
the atmosphere. The atmospheric composition is assumed to be solar, but the only elements 
that are explicitly tracked are H, He, C, and O. Line opacities of CO, CH$_4$ and H$_2$O are included, as well 
as collision induced absorption of H$_2$-H$_2$ \citep[e.g.,][]{bor90, bor02} and H$_2$-He 
\citep{jor00}. The molecular mixing ratios are computed based on the approximate 
method given in the Appendix of \citet{bur99}. The code adopts
the water line wavelengths and strengths from \citet{par97}, the CO lines are from \citet{goo94}
and the CH$_4$ lines are from the High-resolution Transmission Molecular Absorption Database 
\citep[HITRAN,][]{rot98}. The line opacities are computed explicitly 
using Voigt profile line shapes with pressure broadening coefficient of 
$\rm 0.1\,cm^{-1}\,atm^{-1}$. If the water lines are considered individually, 
since there are hundreds of millions of them and many overlap at a given wavelength, 
the model calculations become computationally expensive. To mitigate
this issue, we bin the water lines, to reduce the number of them that needs to 
be included at a given wavelength. The individual water line strengths are dependent on 
temperature, and so is the strength of the binned lines. Therefore, we compute the binned
line strengths for six different temperatures between $300$ and $3000$\,K, and we approximate the 
log of the total strength of the binned lines as a quadratic function of the logarithm of the 
temperature. Since the quadratic is used in log-log space, it is very well able to capture 
the variation of the total binned line strength with temperature. To confirm this, we 
have inspected visually a number of these fits.
The CO and CH$_4$ lines are less numerous and do not require binning. 

The \citet{ric03} code does not incorporate hazes or clouds. Hazes have been recently 
observed in the UV dayside spectrum of HD~189733b \citep{eva13}, but appear to lose 
importance towards visible wavelengths. Hazes also appear to play 
an important role in the formation of the transmission spectrum of HD~189733b (which 
is measured near the terminator, not on the day side, like the emission spectrum) 
at wavelengths below $\sim2.5\,\mu$m \citep[e.g.,][]{pon08, gib12}. \citet{gib12} suggest the existence of a 
Rayleigh-scattering haze in the planet's atmosphere, but this would be expected 
to become transparent at longer wavelengths. In addition, the light paths through 
the atmosphere for the emission spectrum are near vertical, much shorter than the 
planet-grazing light paths observed during transit spectroscopy. Hence, small amounts
of atmospheric haze will have a much smaller cumulative effect on the dayside emission
spectrum than in the transmission spectrum of the planet.  Thus, it is unlikely that
hazes are essential for modeling the infrared dayside emission spectrum of the planet.

In order to calculate the model star-planet contrast, or secondary eclipse depth, 
we adopt a Kurucz model for the host star \citep{kur79} and use its predicted 
brightness at a given wavelength to divide the model planet brightness. This yields 
the model eclipse depth at this wavelength. The most important input for the 
\citet{ric03} model is the pressure-temperature (P-T) and we calculate the model 
emission spectrum for several P-T profiles. 

As \citet{bur14} argues, many of the exoplanet characterization results to date 
are fragile and possibly misleading. Thus, we compare our results on the well-studied 
HD~189733b to models with isothermal, gray and full non-gray hydrostatic equilibrium 
atmospheres in order to test what we know truly robustly. Eliminating the simplest 
models reliably would ensure that we do not infer more 
atmospheric information based on a model than what is justified by the data. 

\subsection{Isothermal Model Atmosphere}
\begin{figure*}
  \epsscale{0.7}
  \plotone{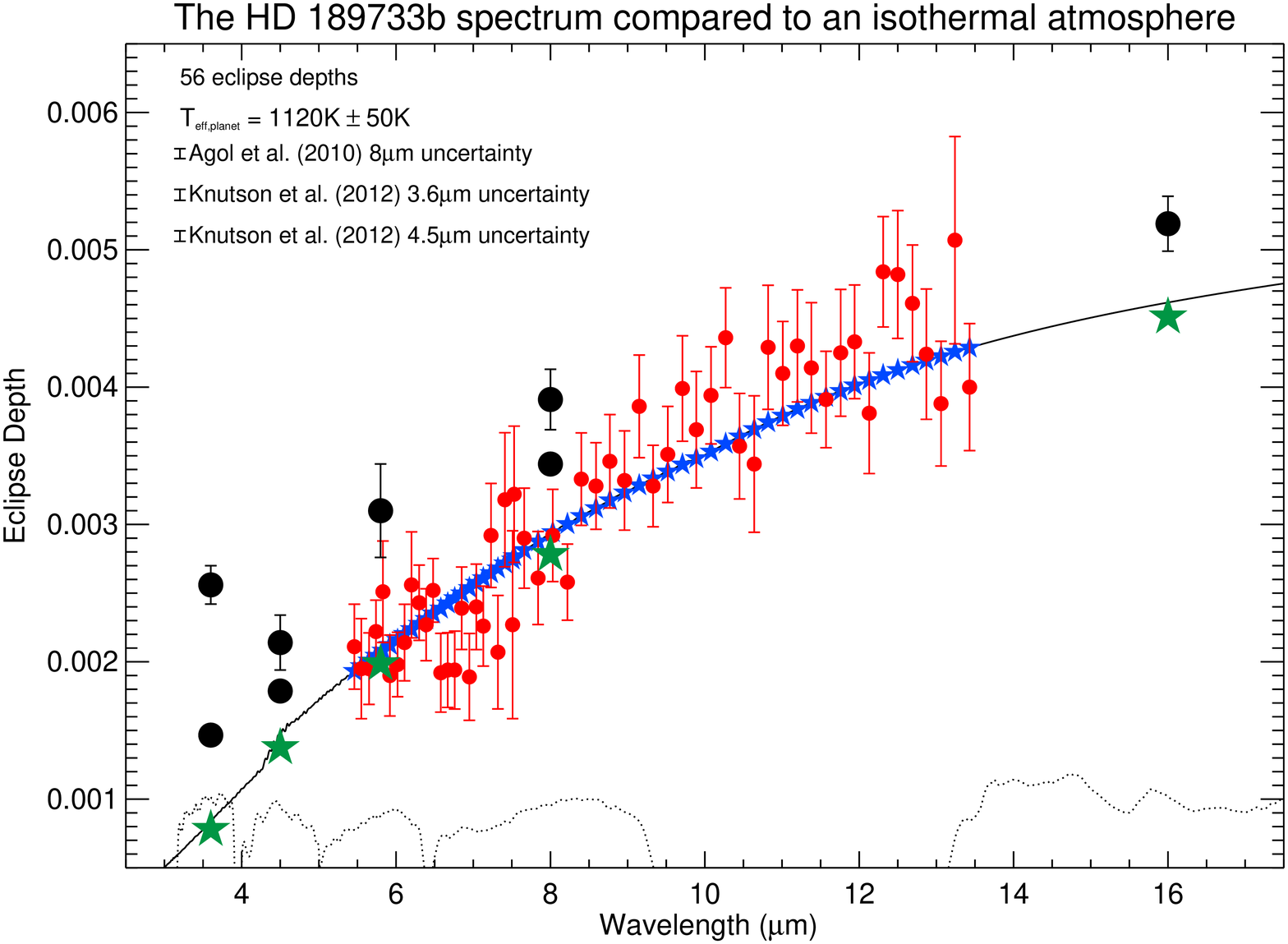}
  \caption{We compare the observed planet-to-star contrast of HD~189733 (large red points) 
    in 56 wavelengths with a prediction based on a blackbody planet, i.e. isothermal atmosphere (solid 
    black line). The broad band secondary eclipse depths measured by \citet[black circles with error bars]{cha08}, 
    \citet{knu12} and \citet[black circles with no error bars at 3.6, 4.5 and 8\,$\mu$m, respectively]{agol10} 
    using {\it Spitzer} IRAC and MIPS are also plotted. The \citet{agol10} measurement at $8\,\mu$m represents 
    an average of six separate eclipse depth measurements. The size of the uncertainties of the 
    \citet{agol10} and \citet{knu12} results is indicated in the upper left for 
    clarity. We show the band integrated contrasts 
    predicted for a blackbody planet for the IRS data (blue stars) and the IRAC and 
    MIPS band passes (green stars). The black dotted lines indicate the transmission functions 
    of {\it Spitzer} IRAC and MIPS. We indicate the effective temperature of the planet
    assuming an isothermal atmosphere in the upper left, with uncertainties derived from the 
    $\chi^2$ distribution and based only on the spectroscopic data points. 
    For this fit, the minimum $\chi^2_{red} = 1.02$. This value does not include the photometric
    points. The only degree of freedom here is the planetary temperature. 
  }
  \label{fig:bbspec}
\end{figure*}

The simplest possible model atmospheric P-T profile is isothermal, which produces
a blackbody emission spectrum. If the measured eclipse depths for this planet
are consistent with blackbody, this would be a strong indication that we can extract
very little information about this planet's atmosphere from our data. A comparison 
between a blackbody planet and our derived spectrum is presented in \autoref{fig:bbspec}.

While for wavelengths of $\lambda \gtrsim 8\,\mu$m the isothermal atmosphere prediction 
matches the observations, it cannot explain well the ``bump''-like feature near 
$6.3\,\mu$m, which is likely caused by water {\it absorption}. We examine this feature in detail in 
Section~\ref{sec:imp}. The shape of the \citet{cha08} photometric spectral energy 
distribution (SED) is also inconsistent with a black body planet, especially at 3.6\,$\mu$m. 
However, this photometric study did not have the benefit of modern analysis techniques. 
In addition, more recent measurements by \citet{knu12} find shallower eclipse depths at 
3.6 and 4.5\,$\mu$m, inconsistent with the corresponding \citet{cha08} measurements. 
Therefore, photometry alone cannot be solely used to rule out the isothermal model. 
 
\subsection{Gray Pressure-Temperature Atmospheric Profile}
\label{sec:gray}
\begin{figure}
  \epsscale{0.7}
  \plotone{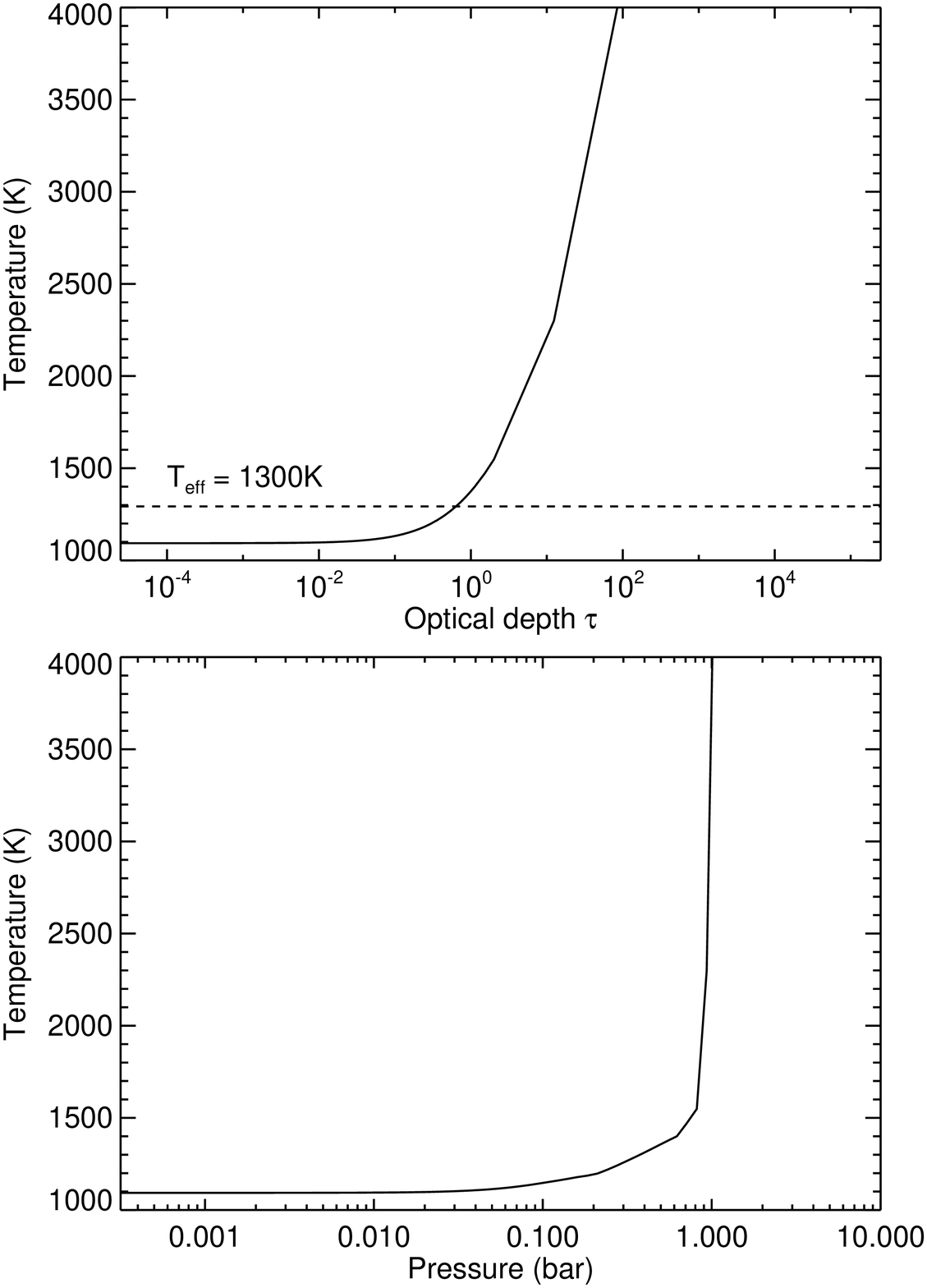}
  \caption{ 
    The calculated temperature as a function of optical depth (upper panel) and
    pressure (lower panel), assuming a gray atmosphere with adopted Rosseland mean opacities
    from \citet{fre08}. The dashed line indicates the effective temperature of the planet, 
    $T_{eff}$, at $\tau=2/3$.}
  \label{fig:graytp}
\end{figure}

\begin{figure*}
  \epsscale{0.7}
  \plotone{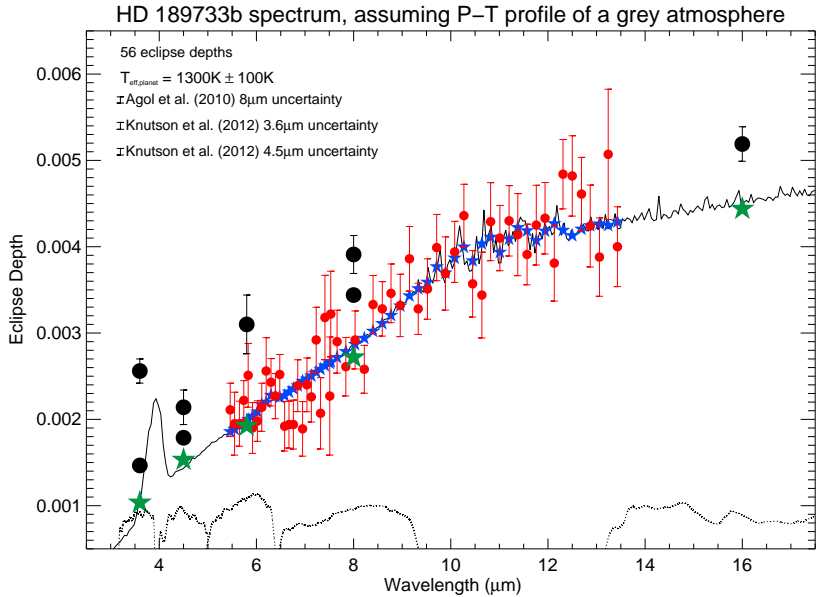}
  \caption{
    Similar to \autoref{fig:bbspec}, but here we show a \citet{ric03} model 
    based on the gray P-T profile computed in Section~\ref{sec:gray} (solid black line) compared to the
    observed results -- large red points with error bars (spectroscopy) and black circles (photometry). 
    The band integrated contrasts 
    predicted for a gray atmosphere are represented by blue stars (spectroscopy) and 
    green stars (photometry). As in \autoref{fig:bbspec}, the effective temperature of the planet, $T_{eff}$, 
    is shown in the upper left, with uncertainties based on the 
    $\chi^2$ distribution, taking into account only the spectroscopic data. 
    A gray atmosphere P-T profile is ruled out
    because it fails to account for the $6.3\,\mu$m bump and for the \citet{cha08} $3.6\,\mu$m eclipse
    depth. The minimum reduced $\chi^2$ value for this fit, assuming that only 
    $T_{eff}$ is free, is $\chi^2_{red} = 0.77$. Again, we only used the spectroscopic points 
    for the fit and the $\chi^2_{red}$ calculation. As in the black body case, here there is only 
    one degree of freedom, $T_{eff}$. 
  }
  \label{fig:gaspec}
\end{figure*}

A gray atmosphere, where the gas opacity is equal at all 
wavelengths, has a more complex P-T profile than an isothermal atmosphere 
\citep[for a detailed discussion of gray atmospheres see, e.g.,][]{rut03}. We adopt the 
Rosseland mean opacities for Solar metallicity given by \citet{fre08} for ultracool brown dwarfs and
exoplanets for temperatures between 75 and 4000$\,K$, and interpolate between the 
values provided in their tables as necessary. We begin with an ``initial guess'' P-T
profile (isothermal one) and use it to estimate the run of Rosseland opacity with pressure. We then 
calculate the optical depth and, hence, temperature as a function of pressure. Using 
this, we re-estimate the Rosseland opacity as a function of pressure and repeat the 
process until the run of temperatures with pressure  converges. The effective temperature of the planet, $T_{eff}$, is 
given by the temperature at optical depth, $\tau=2/3$ and this is a model input parameter, 
represented by the temperature of the initial isothermal profile guess. Its value sets
absolute values of the P-T profile, as opposed to the profile's shape, which is set 
by the gray atmosphere. We experiment with plugging the gray P-T profile for a range of 
values for $T_{eff}$ in the \citet{ric03} radiative transfer code, where the atmosphere 
is {\it not} assumed to be gray and examine the resulting planetary emission spectra. 
Using $\chi^2$ minimization, we find that $T_{eff} = 1300\,K \pm 100\,K$ yields an emergent spectrum 
closest to the observed one. The uncertainties on this quantity are based on the $\chi^2$
distribution of the spectroscopic data, assuming that only $T_{eff}$ is a free parameter.
Here, we exclude the photometric data points, since it is difficult to estimate 
their systematic offset and thus to assign them appropriate weights. We show the results from this calculation in 
Figures~\autoref{fig:graytp} and \autoref{fig:gaspec}. As with the isothermal model, a 
gray atmosphere P-T profile appears to be unlikely based on the shape of the \citet{cha08} broadband SED, 
especially the $3.6\,\mu$m eclipse depth, despite lack of updated reduction of these data.
However, a gray atmosphere appears to be consistent with the more recent \citet{knu12}
photometric data points at 3.6 and 4.5$\,\mu$m. 
As with the isothermal model, the gray atmosphere fails to account for the secondary 
eclipse spectrum near the $6.3\,\mu$m ``bump,'' discussed in Section~\ref{sec:imp}.

\subsection{The Burrows Atmosphere Model}
\label{sec:bur}

\begin{figure}
  \epsscale{0.7}
  \plotone{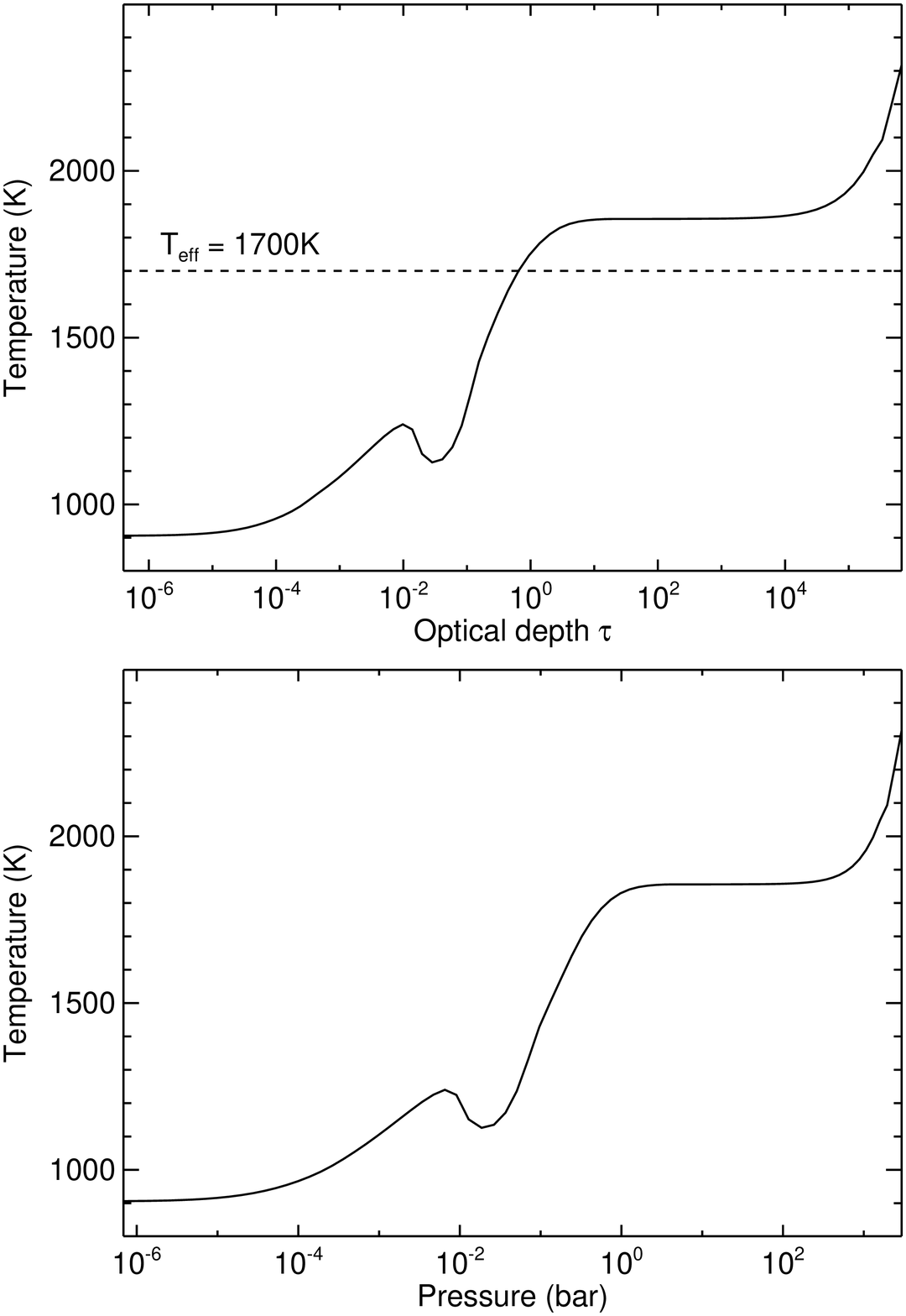}
  \caption{ 
    The temperature as a function of optical depth based on the Rosseland opacity (upper panel) and
    pressure (lower panel) used to calculate the Burrows model similar to the one adopted by \citet{gri08}.
    As in \autoref{fig:graytp}, the dashed line indicates the effective temperature of the planet, 
    $T_{eff}$. Here, the high-altitude absorber is assumed to be relatively unimportant for the emission spectrum, 
    with $\kappa_{abs}=0.02\,{\rm cm^2g^{-1}}$, but it still causes a small temperature inversion 
    near pressure of $0.01\,$bar.
  }
  \label{fig:burtp}
\end{figure}

\begin{figure*}
  \epsscale{0.7}
  \plotone{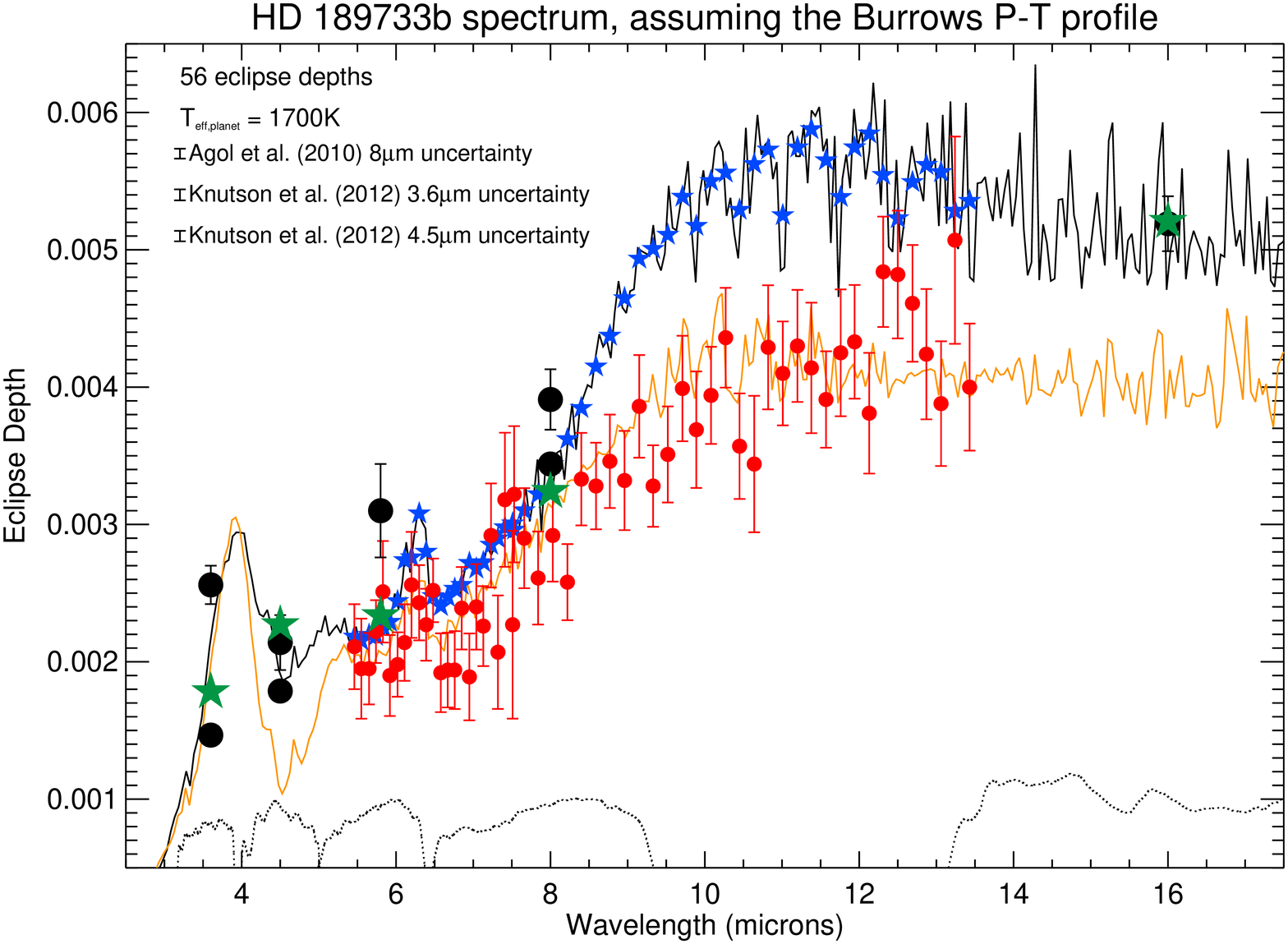}
  \caption{
    Similar to Figures~\autoref{fig:bbspec} and \autoref{fig:gaspec}, but here we compare our observed 
    planet-star contrast (large red points) to two models produced based on the P-T profile shown in 
    \autoref{fig:burtp}. These models a prohibitively computationally intensive to calculate, 
    making statistical fitting impractical. Therefore, we do not fit them to the data, 
    but we plot them with the observed data for 
    comparative purposes. The planetary emission model similar to the one adopted by 
    \citet{gri08} is calculated based on the studies by \citet[][solid orange line]{bur07,bur08}. 
    The black line is a simpler \citet{ric03} model based on the \citet{gri08} P-T profile
    that we calculate. As before, the star symbols represent a band-integrated version of the this model. We do not
    claim that any of these two models is preferred, but they give a better match to 
    the shape of the planetary spectrum, compared to the isothermal and gray models, 
    even though they both predict deeper eclipses than observed. For the better matching 
    \citet{bur07,bur08} model, we calculate the non-optimized $\chi^2 = 406.8$, based on the 
    56 IRS eclipse depths (this value does not take the Spitzer IRAC photometry into account). While 
    there are numerous input parameters for the Burrows models 
    they are all kept fixed here. Thus, we do not calculate the reduced $\chi^2$ value. 
  }
  \label{fig:burspec}
\end{figure*}

In their study, \citet{gri08} compare their results to a model developed by \citet{bur07,bur08} that 
adopts the chemical equilibrium and opacities information computed by \citet{bur99} and \citet{sha07}. 
The model relies on fully non-gray radiative opacities with layer-by-layer radiative equilibrium
and chemical equilibrium using an extensive set of molecular and atomic abundances. The model also includes a parametrized generic
stratospheric flux absorber that can cause a temperature inversion, and a 
parametrized day-side-to-night-side heat transfer efficiency. We compare the Burrows model adopted 
by \citet{gri08} to our measurements. We also use the Burrows-derived P-T profile 
as input to the \citet{ric03} radiative transfer code, and also compare the result to our observations. 
The Burrows model presented here is similar to the one shown in \citet{gri08} and 
has a heat redistribution parameter $P_n=0.1$ implying that
only 10\% of the heat absorbed on the planet's day side is transferred to the night side (the maximum is
$P_n=0.5$, or 50\%). The absorption coefficient of the hypothetical unknown high-altitude absorber is set
to $\kappa_{abs}=0.020\,{\rm cm^2g^{-1}}$, meaning that the planet is assumed to have no or negligible 
stratosphere and relatively inefficient heat transfer to its night side. Later studies have also found no 
evidence for inversion \citep{mad09,swa09b}, and the planet is typically considered to have a non-inverted
atmosphere \citep[e.g.,][]{knu10}. The Burrows model cannot be ruled 
out based on our analysis. We present the P-T pressure for this
model in \autoref{fig:burtp} and a comparison between the emission from the Burrows atmosphere to 
our observed eclipse depths in \autoref{fig:burspec}.

\subsection{Discussion of Results and Implications for the Atmosphere}
\label{sec:imp}

\begin{figure}
  \epsscale{1.1}
  \plotone{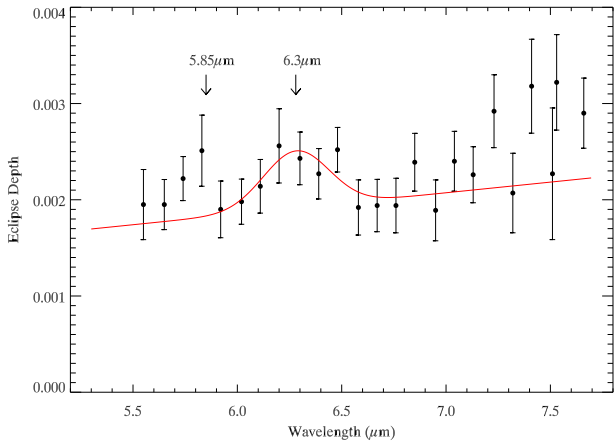}
  \caption{ 
    A section of the observed secondary eclipse spectrum of HD~189733b near the $6\,\mu$m water feature. 
    This apparent ``bump'' in the spectrum is caused by an opacity minimum near $6.27\,\mu$m between two 
    {\it absorption} features of water on either side (see text for details). 
    We have over-plotted a Gaussian curve centered at $6.32\,\mu$m with width of $0.15\,\mu$m. 
    We also mark the tenuous unidentified emission bump near $5.9\,\mu$m \citep{gri08}, which we consider 
    to be most likely a noise artifact.  
  }
  \label{fig:6um}
\end{figure}

\citet{gri08} describe a ``bump'' in their spectrum near $6.2\,\mu$m that they attribute to the
opacity minimum between the P and R branches of the $\nu_2$ band at $6.27\,\mu$m (the
fundamental vibrational bending mode of water).
These authors also notice a very tenuous rise in emission at $5.9\,\mu$m that 
they are unable to identify.
These features are preserved after the inclusion of the previously never
analyzed observations in this study (\autoref{fig:6um}). Even though the flux near 
$5.9\,\mu$m still appears higher than that of its neighbors, the uncertainties are too 
large to be able to claim even a tentative detection. 

In order to test the robustness of the detection of the $6\,\mu$m water feature, we focus on the
$5.9-7.0\,\mu$m range of the results. We fit a Gaussian function added to a sloped line and measure an 
amplitude of the feature of $0.00058$ in units of contrast and a wavelength of the
maximum at $6.28\,\mu$m, with a Gaussian width, $\sigma=0.15\,\mu$m. We run a Monte
Carlo simulation of 10,000 spectra in this range, by drawing random numbers from 
Gaussian distributions with means equal to the fitted straight line but with the 
Gaussian feature removed and with widths equal to the observed eclipse depth uncertainties. 
We fit every simulated spectrum with a Gaussian in the same way we fit the observed water 
feature and find that only $\sim 3\%$ of the simulated data sets exhibit peaks similar 
to that of the real spectrum -- with amplitudes greater than $0.0001$ and widths 
between $0.1$ and $0.2\,\mu$m. 

Therefore, we rule out a smooth spectrum for this wavelength range
and thus reject the isothermal model for the atmosphere at the $97\%$ level, based solely on 
the IRS data. Our analysis is in agreement with the conclusions of \citep{gri08}, who detect the $6\,\mu$m water 
feature at the $95\%$ level. 

The shape of the gray atmosphere model is similar to the shape to the blackbody spectrum at 
$5-7\,\mu$m wavelength range, especially at low wavelength resolution, 
as in our data (\autoref{fig:gaspec}). Therefore, the gray atmosphere model can be 
rejected at the 97\% level with the same argument as the isothermal model -- it is in disagreement with the detection 
of the $6\,\mu$m feature. 

The Burrows and \citet{ric03} models based on the P-T profile similar to the one 
adopted in \citet{gri08} provide the best match to the
shape of the observed spectrum, especially near the $6\,\mu$m water feature. However, both of these predict 
deeper eclipses than the ones we observe. This makes the Burrows-based models more 
difficult to reject outright, even though they 
do not provide a perfect match for the measurements. 

For completeness, we compare the $\chi^2$ statistics produced by the three models
(excluding the broad band eclipses, whose weights in the calculation would be non-trivial). 
The isothermal atmosphere model produces $\chi^2 = 54.9$, larger than the value of $\chi^2 = 41.4$
produced by the gray atmosphere model (our spectrum consists of 56 points). 
For the full Burrows radiative 
transfer model, $\chi^2 = 406.8$. While the $\chi^2$ values strongly favor the 
simpler models, they are a poor match to the observed data near $6\,\mu$m, and 
we have not minimized the $\chi^2$ parameter for the Burrows model as this is impractical.
More importantly, the $\chi^2$ statistic is very sensitive to the absolute eclipse depth values 
compared to the model, so it is optimal for testing the absolute eclipse depths, 
not the shape of the spectrum, on which we have focused here. Since a true fit is 
not feasible, we attempt applying absolute shifts to the data and comparing it to the
Burrows model, until the $\chi^2$ is minimized. We find that $\chi^2 = 160.2$. 
However, drawing any conclusions out of this (e.g., that the model is rejected) is 
difficult, because a true model fit with realistic free parameters 
is needed in order to show this definitely.

While the combination of photometry and spectroscopy proves the 
isothermal or gray atmosphere explanation for the observations unlikely, the rejection 
hinges on data within a small wavelength range. This underscores the necessity of large wavelength 
coverage and high spectral resolution required to make definitive measurements of the properties 
of exoplanet atmospheres, as suggested by the model retrieval efforts by \citet{lee12,lin12,lin13}.
While the Burrows model cannot be rejected outright, it is possible that additional data in other wavelengths
may show that it is an insufficient explanation of the atmosphere of this hot Jupiter. 

\section{Conclusion}
\label{sec:con}
Our analysis of the {\it Spitzer}/IRS secondary eclipse time series spectroscopy of HD~189733b
has largely confirmed the results from the original \citet{gri08} study, despite the fact that
it did not include many of the observations available to us. While we confirm the detection of the 
$6\,\mu$m water feature, we find that broad wavelength coverage and high spectral resolution 
are essential for the studies of exoplanet atmospheres, since we are only able to reject the
simplistic isothermal and gray atmosphere models based on only several out of a total of $\sim60$ eclipse 
depth measurements. The fact that our results are in broad agreement with the \citet{gri08}
is an important confirmation that many of the systematics correction techniques the exoplanet community has 
been employing for {\it Spitzer} observations are robust and can be applied to the next generation 
of space-based exoplanet studies as necessary. The offset between the \citet{cha08} photometry 
and the updated spectroscopy in this work, as well as the photometry in \citet{knu12}, 
should be examined further, in particular with
a consistent updated analysis of all secondary eclipse photometric data sets available for this planet. 
The offset is likely related to the choice of ramp correction function. 
HD~189733b is the exoplanet with the most thoroughly studied 
atmosphere to date and our results enhance this achievement, despite the
fact that plenty of questions about its atmosphere's structure and composition remain open.  
These questions could be answered by future studies in the context of new and archival 
observations. 

\acknowledgements
We thank the {\it Spitzer} Science Center HelpDesk and Sean Carey for their valuable assistance in 
exploring alternatives for the correction of the systematic effects caused 
by the telescope pointing jitter. We thank David Charbonneau and Jonathan Fortney 
for valuable discussions. 
We thank the referee for the detailed review and many thoughtful suggestions. 
This work is based on observations made with the Spitzer Space Telescope, 
obtained from the NASA/IPAC Infrared Science Archive, both of which are 
operated by the Jet Propulsion Laboratory, California Institute of Technology 
under a contract with the National Aeronautics and Space Administration.
Support for this work was provided by NASA through an 
award issued by JPL/Caltech. This research has made use of NASA's Astrophysics Data System.
This research has made use of the Exoplanet 
Orbit Database and the Exoplanet Data Explorer at \url{exoplanets.org}.


\newpage 
\begin{thebibliography}{}
\bibitem[Agol et al.(2010)]{agol10}
Agol, E., Cowan, N. B., Knutson, H. A., Deming, D., Steffen, J. H.,
Henry, G. W., \& Charbonneau, D. 2010, \apj, 721, 1861

\bibitem[Barstow et al.(2013)]{bar13} 
Barstow, J.~K., Aigrain, S., Irwin, P.~G.~J., Fletcher, L.~N., 
\& Lee, J.-M.\ 2013, \mnras, 434, 2616 

\bibitem[Beerer et al.(2011)]{bee11}
Beerer, I.~M., et al. 2011, \apj, 727, 23

\bibitem[Birkby et al.(2013)]{bir13} 
Birkby, J.~L., Ce Kok, R.~J., Brogi, M., et al.\ 2013, \mnras, 436, L35 

\bibitem[Borysow \& Frommhold(1990)]{bor90} 
Borysow, A., \& Frommhold, L.\ 1990, ApJL, 348, L41 

\bibitem[Borysow(2002)]{bor02} 
Borysow, A.\ 2002, A\&A, 390, 779 

\bibitem[Burrows (2014)]{bur14} 
Burrows, A.\ 2014, PNAS, doi: 10.1073/pnas.1304208111, accepted

\bibitem[Burrows \& Sharp(1999)]{bur99} 
Burrows, A., \& Sharp, C.~M.\ 1999, \apj, 512, 843 

\bibitem[Burrows et al.(2007)]{bur07}
Burrows, A., Hubeny, I., Budaj, J., Knutson, H.~A., 
\& Charbonneau, D. 2007, \apjl, 668, L171 

\bibitem[Burrows et al.(2008)]{bur08}
Burrows, A., Budaj, J., \& Hubeny, I. 2008, \apj, 678, 1436 

\bibitem[Charbonneau et al.(2005)]{cha05}
Charbonneau, D., et al. 2005, ApJ, 626, 523

\bibitem[Charbonneau et al.(2008)]{cha08} 
Charbonneau, D., Knutson, H.~A., Barman, T., Allen, L.~E., 
Mayor, M., Megeath, S. T., Queloz, D., \& Udry, S. 2008, \apj, 686, 1341

\bibitem[Correia \& Laskar(2010)]{cor10}
{Correia}, A.~C.~M. \& {Laskar}, J. 2010, in Exoplanets, ed. S.~Seager,
(Tucson: Univ. Arizona Press), 239

\bibitem[Cowan \& Agol(2011)]{cow11} 
{Cowan}, N.~B., \& {Agol}, E. 2011, \apj, 729, 54

\bibitem[Crossfield et al.(2012)]{cro12} 
Crossfield, I.~J.~M., Hansen, B.~M.~S., \& Barman, T.\ 2012, \apj, 746, 46 

\bibitem[Deming et al.(2005)]{dem05}
{Deming}, D., {Seager}, S., {Richardson}, \& 
L.~J., {Harrington}, J. 2005, Nature, 434, 740

\bibitem[Deming et al.(2006)]{dem06} 
Deming, D., Harrington, J., Seager, S., \& Richardson, L.~J.\ 2006, \apj, 644, 560 

\bibitem[Deming et al.(2011)]{dem11} 
{Deming}, D., et al. 2011, \apj, 726, 95

\bibitem[D{\'e}sert et al.(2011)]{des11} 
D{\'e}sert, J.-M., Charbonneau, D., Fortney, J.~J., et al.\ 2011, \apjs, 197, 11 

\bibitem[Eastman et al.(2010)]{eas10} 
{Eastman}, J., {Siverd}, R. \& {Gaudi}, B.~S. 2010, \pasp, 122, 935

\bibitem[Evans et al.(2013)]{eva13} 
Evans, T.~M., Pont, F., Sing, D.~K., et al.\ 2013, \apjl, 772, L16 

\bibitem[Ford(2005)]{ford05} 
Ford, E.~B.\ 2005, \aj, 129, 1706 

\bibitem[Ford(2006)]{ford06} 
Ford, E.~B.\ 2006, \apj, 642, 505 

\bibitem[Fortney et al.(2006)]{for06}
Fortney, J. J., Saumon, D., Marley, M. S., Lodders, K., \&
Freedman, R. S. 2006, ApJ, 642, 495

\bibitem[Fortney et al.(2008)]{for08}
Fortney, J. J., Lodders, K., Marley, M. S., \& Freedman, R. S. 2008, ApJ, 678, 1419

\bibitem[Freedman et al.(2008)]{fre08} Freedman, R.~S., 
Marley, M.~S., \& Lodders, K.\ 2008, ApJS, 174, 504 

\bibitem[Gibson et al.(2012)]{gib12} 
Gibson, N.~P., Aigrain, S., Pont, F., et al.\ 2012, MNRAS, 422, 753 

\bibitem[Gillon et al.(2007)]{gil07} 
Gillon, M., Demory, B.-O., Barman, T., et al.\ 2007, \aap, 471, L51 

\bibitem[Goorvitch(1994)]{goo94} 
Goorvitch, D.\ 1994, \apjs, 95, 535 

\bibitem[Grillmair et al.(2007)]{gri07} Grillmair, C.~J., 
Charbonneau, D., et al.\ 2007, \apjl, 658, L115 

\bibitem[Grillmair et al.(2008)]{gri08} 
Grillmair, C.~J., et al.\ 2008, \nat, 456, 767 

\bibitem[Henry \& Winn(2008)]{hen08} 
Henry, G.~W., \& Winn, J.~N.\ 2008, \aj, 135, 68 

\bibitem[Horne(1986)]{hor86} 
Horne, K.\ 1986, PASP, 98, 609 

\bibitem[Houck et al.(2004)]{hou04} 
Houck, J.~R., Roellig, T.~L., van Cleve, J., et al.\ 2004, ApJS, 154, 18 

\bibitem[Hubeny et al.(2003)]{hub03}
Hubeny, I., Burrows, A., \& Sudarsky, D. 2003, ApJ, 594, 1011

\bibitem[Jackson et al.(2008)]{jac08} 
Jackson, B., Greenberg, R., \& Barnes, R.\ 2008, \apj, 678, 1396 

\bibitem[J{\o}rgensen et al.(2000)]{jor00} 
J{\o}rgensen, U.~G., Hammer, D., Borysow, A., \& Falkesgaard, J.\ 2000, A\&A, 361, 283 

\bibitem[Knutson et al.(2007)]{knu07}
Knutson, H.~A, et al. 2007, \nat, 447, 183

\bibitem[Knutson et al.(2008)]{knu08} 
Knutson, H.~A., Charbonneau, D., Allen, L.~E., 
Burrows, A., \& Megeath, S.~T. 2008, \apj, 673, 526

\bibitem[Knutson et al.(2010)]{knu10} 
Knutson, H.~A., Howard, A.~W., \& Isaacson, H. 2010, \apj, 720, 1569

\bibitem[Knutson et al.(2012)]{knu12} 
  Knutson, H.~A., Lewis, N., Fortney, J.~J., et al.\ 2012, \apj, 754, 22 

\bibitem[Kurucz(1979)]{kur79}
Kurucz, R. L. 1979, ApJS, 40, 1

\bibitem[Lee et al.(2012)]{lee12} 
Lee, J.-M., Fletcher, L.~N., \& Irwin, P.~G.~J.\ 2012, \mnras, 420, 170 

\bibitem[Lewis et al.(2013)]{lew13} 
Lewis, N.~K., Knutson, H.~A., Showman, A.~P., et al.\ 2013, \apj, 766, 95 

\bibitem[Line et al.(2012)]{lin12} 
Line, M.~R., Zhang, X., Vasisht, G., et al.\ 2012, \apj, 749, 93 

\bibitem[Line et al.(2013)]{lin13} 
Line, M.~R., Wolf, A.~S., Zhang, X., et al.\ 2013, \apj, 775, 137 

\bibitem[Machalek et al.(2009)]{mac09}
Machalek, P., McCullough, P.~R., Burrows, A., Burke, C.~J., 
Hora, J.~L., \& Johns-Krull, C.~M. 2009, \apj, 701, 514

\bibitem[Madhusudhan \& Seager(2009)]{mad09} 
Madhusudhan, N., \& Seager, S.\ 2009, ApJ, 707, 24 

\bibitem[Mandel \& Agol(2002)]{man02} 
Mandel, K., \& Agol, E.\ 2002, \apjl, 580, L171 

\bibitem[Markwardt(2009)]{mar09} 
Markwardt, C.~B.\ 2009, Astronomical Data Analysis Software and Systems XVIII, 411, 251

\bibitem[Parmentier et al.(2013)]{par13}
Parmentier, V., Showman, A. P., \& Lian, Y. 2013, A\&A, submitted

\bibitem[Partridge \& Schwenke(1997)]{par97} 
Partridge, H., \& Schwenke, D.~W.\ 1997, J. Chem. Phys., 106, 4618 

\bibitem[Pont et al.(2007)]{pon07} 
Pont, F., Gilliland, R.~L., Moutou, C., et al.\ 2007, \aap, 476, 1347 

\bibitem[Pont et al.(2008)]{pon08} 
Pont, F., Knutson, H., Gilliland, R.~L., Moutou, C., \& Charbonneau, D.\ 2008, MNRAS, 385, 109 

\bibitem[Richardson et al.(2003)]{ric03} 
Richardson, L.~J., Deming, D., \& Seager, S.\ 2003, ApJ, 597, 581 

\bibitem[Richardson et al.(2007)]{ric07} 
Richardson, L.~J., Deming, D., Horning, K., Seager, S., \& Harrington, J.\ 2007, \nat, 445, 892 

\bibitem[Rothman et al.(1998)]{rot98} 
Rothman, L.~S., Rinsland, C.~P., Goldman, A., et al.\ 1998, \jqsrt, 60, 665 

\bibitem[Rutten(2003)]{rut03} 
Rutten, R.~J.\ 2003, Radiative Transfer in Stellar Atmospheres, by 
Robert J.~Rutten.~Lecture Notes Utrecht University

\bibitem[Schwarz (1978)]{sch1978} 
Schwarz, G.\ 1978, Ann. Statist., 6, 461 

\bibitem[Sharp \& Burrows(2007)]{sha07} 
Sharp, C.~M., \& Burrows, A.\ 2007, \apjs, 168, 140 

\bibitem[Stevenson et al.(2010)]{ste10} 
Stevenson, K.~B., et al.\ 2010, \nat, 464, 1161 

\bibitem[Swain et al.(2008b)]{swa08a} 
Swain, M.~R., Bouwman, J., Akeson, R.~L., Lawler, S., \& Beichman, C.~A.\ 2008a, \apj, 674, 482 

\bibitem[Swain et al.(2008a)]{swa08b} 
Swain, M.~R., Vasisht, G., \& Tinetti, G.\ 2008b, \nat, 452, 329 

\bibitem[Swain et al.(2009a)]{swa09a} 
Swain, M.~R., et al.\ 2009a, \apj, 704, 1616 

\bibitem[Swain et al.(2009b)]{swa09b} 
Swain, M.~R., Vasisht, G., Tinetti, G., et al.\ 2009b, \apjl, 690, L114 

\bibitem[Tinetti et al.(2007)]{tin07} Tinetti, G., 
Vidal-Madjar, A., Liang, M.-C., et al.\ 2007, \nat, 448, 169 

\bibitem[Todorov et al.(2010)]{tod10}
Todorov, K.~O., Deming, D., Harrington, J., Stevenson, K. B., Bowman, W. C., 
Nymeyer, S., Fortney, J. J., \& Bakos, G. A. 2010, \apj, 708, 498

\bibitem[Todorov et al.(2012)]{tod12} Todorov, K.~O., Deming, 
D., Knutson, H.~A., et al.\ 2012, \apj, 746, 111 

\bibitem[Todorov et al.(2013)]{tod13} Todorov, K.~O., Deming, 
D., Knutson, H.~A., et al.\ 2013, \apj, 770, 102 

\bibitem[Triaud et al.(2009)]{tri09} 
Triaud, A.~H.~M.~J., Queloz, D., Bouchy, F., et al.\ 2009, \aap, 506, 377 

\bibitem[Zahnle et al.(2009)]{zah09}
Zahnle, K., Marley, M. S., Freedman, R. S., Lodders, K., \& Fortney, J. J. 2009, ApJ, 701, L20

\end{thebibliography}
\end{document}